\documentclass[acmsmall]{acmart}
\usepackage{array}

\usepackage{multirow}
\newcommand{\revised}[1]{#1}

\AtBeginDocument{%
  }

\acmSubmissionID{6433}

\begin{document}

\setcopyright{cc}
\setcctype{by}
\acmJournal{PACMHCI}
\acmYear{2026} \acmVolume{10} \acmNumber{5} \acmArticle{MHCI6433}
\acmMonth{8} \acmDOI{10.1145/3821669}

\title{Exploring Context-aware and LLM-driven Locomotion for Immersive Virtual Reality}

\author{Suleyman Ozdel}
\orcid{0000-0002-3390-6154}
\email{ozdelsuleyman@tum.de}
\affiliation{%
  \institution{Human-Centered Technologies for Learning, Technical University of Munich}
  \city{Munich}
  \country{Germany}}
\affiliation{%
  \institution{Munich Center for Machine Learning}
  \city{Munich}
  \country{Germany}}

\author{Kadir Burak Buldu}
\orcid{0009-0005-7877-9602}
\email{burak.buldu@tum.de}
\affiliation{%
  \institution{Human-Centered Technologies for Learning, Technical University of Munich}
  \city{Munich}
  \country{Germany}}

\author{Enkelejda Kasneci}
\orcid{0000-0003-3146-4484}
\email{enkelejda.kasneci@tum.de}
\affiliation{%
  \institution{Human-Centered Technologies for Learning, Technical University of Munich}
  \city{Munich}
  \country{Germany}}
\affiliation{%
  \institution{Munich Center for Machine Learning}
  \city{Munich}
  \country{Germany}}

\author{Efe Bozkir}
\orcid{0000-0002-4594-4318}
\email{efe.bozkir@tum.de}
\affiliation{%
  \institution{Human-Centered Technologies for Learning, Technical University of Munich}
  \city{Munich}
  \country{Germany}}

\renewcommand{\shortauthors}{Ozdel et al.}

\begin{abstract}
Locomotion shapes the usability and comfort of immersive virtual environments. In particular, hands-free locomotion supports accessibility, yet most speech-based approaches rely on rigid command sets that limit natural and flexible interaction. We introduce a speech-based, context-aware, LLM-driven hands-free locomotion technique that enables free-form natural language navigation. We evaluate our approach against two baselines: controller-based teleportation and fixed-command voice steering. Our evaluation combines eye tracking and standardized questionnaires to assess usability (SUS), presence (IPQ), cybersickness (CSQ-VR), and cognitive load (NASA-TLX), complemented by an SHAP-based analysis. We found no evidence of differences in usability, presence, cognitive load, or cybersickness compared to established methods such as teleportation, while eye-tracking results suggest increased attention and engagement in the LLM-driven condition. In terms of performance, teleportation achieved the shortest task completion times, whereas the two hands-free voice methods (LLM-driven and fixed-command) were comparable. Overall, our results demonstrate the feasibility of LLM-driven hands-free locomotion. This technique is applicable to accessibility use cases such as hands-busy scenarios, and motivates follow-up studies with larger samples and more diverse tasks.
\end{abstract}


\begin{teaserfigure}
  \includegraphics[width=\textwidth]{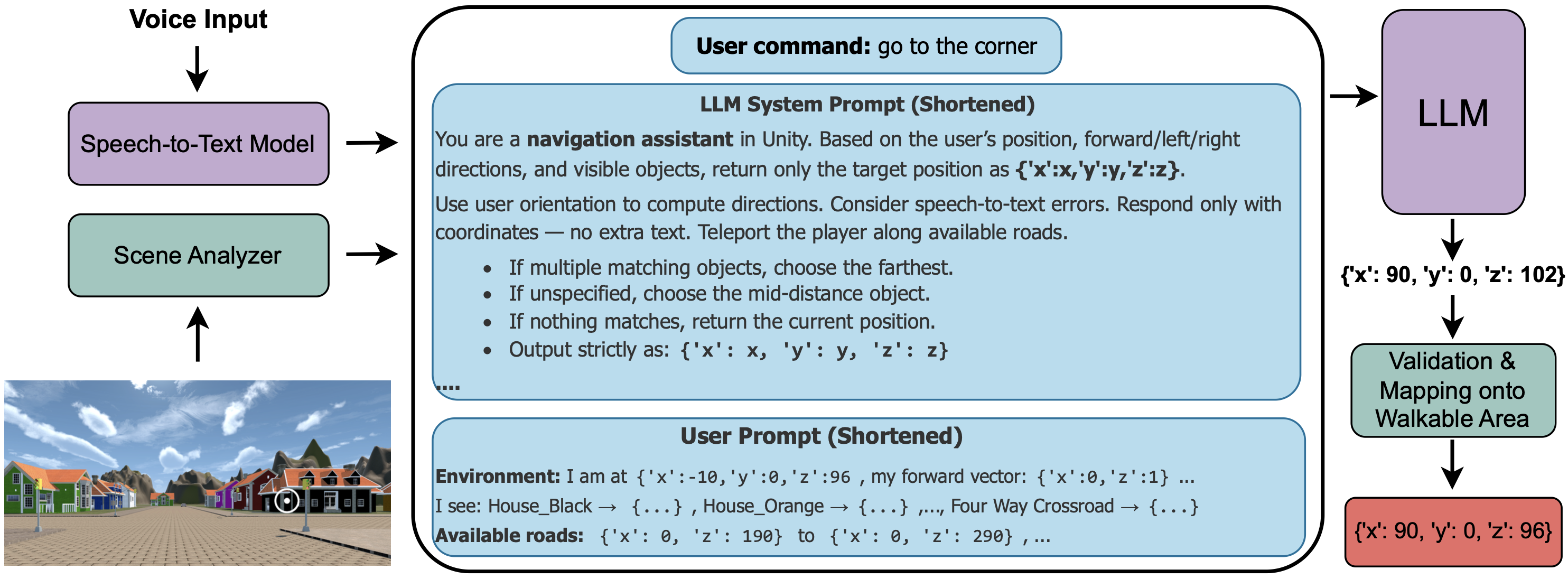}
  \caption{Overview of our LLM-driven locomotion system in VR.}
  \Description{Overview of the LLM-driven voice navigation pipeline in Unity. Voice input is first converted to text via a speech-to-text model, while a scene analyzer extracts the user’s position, orientation, visible objects, and available roads from the environment. These inputs are combined into a structured system and user prompt and passed to the LLM, which acts as a navigation assistant and outputs only target world coordinates \{x,y,z\}. The predicted target is then validated and mapped onto the walkable area, resulting in a final navigable position used to teleport or move the player within the scene.}
  \label{fig:teaser}
\end{teaserfigure}


\maketitle

\section{Introduction}
With the rapid advancement of virtual reality (VR) technologies, VR systems have started to be used widely in various domains and purposes, such as education~\cite{gao2021digital}, entertainment\revised{~\cite{bates1992virtual,ansari2022implementing}}, healthcare~\cite{halbig2022opportunities} and training~\cite{xie2021review}. The quality of interaction in immersive environments becomes especially important in situations where users' hands are occupied, making handheld controllers impractical. Examples include training simulations, field work, or assistive contexts. One of the important \revised{aspects} of user experience is locomotion, which refers to how users move within virtual environments. Locomotion affects many factors including user immersion~\cite{bozgeyikli2019locomotion}, task performance~\cite{coomer2018evaluating}, overall comfort~\cite{hombeck2023tell}, and more importantly cybersickness~\cite{clifton2020effects}. Among various techniques, teleportation by using handheld controllers remains one of the most widely adopted methods~\cite{prithul2021teleportation}.

However, hands-free locomotion is particularly important when users must keep their hands free, such as during multitasking or for applications focusing on accessibility\revised{~\cite{monteiro2021hands,prithul2022evaluation}}. In contexts similar to mobile interaction, speech input is appealing, but it must remain usable despite limitations related to latency, noise, ambiguity, and privacy. To address this, researchers proposed different locomotion techniques with varying input modalities. These modalities includes voice~\cite{ferracani2017natural,calandra2022comparison,hombeck2023tell}, gaze~\cite{christou2017steering}, and gesture~\cite{calandra2019usability,liu2021force}. Among those, voice-based systems provide a natural and intuitive alternative, with techniques that include continuous voice-based steering~\cite{hombeck2023tell} and teleportation methods controlled through verbal destination input~\cite{ferracani2017natural,calandra2022comparison}. Many prior voice-based approaches rely on rule-based grammars, predefined destination names, or fixed command sets~\cite{ferracani2017natural,calandra2022comparison,hombeck2023tell}, which can constrain how users phrase requests and introduce additional learning and disambiguation overhead. These limitations may reduce flexibility and perceived naturalness, especially in dynamic environments, and can negatively affect user satisfaction, immersion, and usability.

Recent advancements in large language models (LLMs), including GPT-family models, offer new opportunities for developing more natural and intelligent voice-based interaction systems in VR~\cite{bozkir2024embedding}. These models can interpret natural language instructions and effectively identify intent in VR, particularly when environmental context is included in the prompt~\cite{skyba2024towards,buldu2025cuify}. Moreover, as they support multiple languages, it is possible to enable multilingual interaction within the same system setup and make virtual environments more generic with minimal effort while supporting user diversity\cite{bozkir2024embedding,buldu2025cuify}. In our prototype and evaluation, we use ChatGPT-4o as the language-understanding component.

In this study, we present an LLM-driven locomotion technique that enables natural language, hands-free navigation in VR, aiming to reduce reliance on fixed command grammars used in many voice-based techniques. Steering-based voice locomotion typically relies on continuous movement, which can cause discomfort and often uses rigid, predefined commands~\cite{hombeck2023tell,mayor2019comparative}. While teleportation-based voice methods reduce motion sickness through instant movements, they still depend on rule-based grammar structures and often involve manual object labeling or predefined mappings for speech recognition and semantic understanding~\cite{ferracani2017natural,calandra2022comparison}. Our approach allows users to express free-form spoken instructions without memorizing fixed commands or grammar structures. The system interprets these instructions in real time using contextual information dynamically extracted from the virtual environment, such as object names, colors, and positions. This reduces the need for handcrafted command rules by allowing both landmark-based requests (e.g., object or place references available in the scene) and relative movement instructions (e.g., distance or direction commands)
, while still relying on the scene information available to the runtime system, as discussed in the limitations.

To evaluate the feasibility of our method, we conducted a user study in a town-like virtual environment, comparing three locomotion techniques: teleportation, voice-based steering through fixed commands, and our LLM-driven locomotion. We collected user feedback through standardized questionnaires and behavioral data using eye tracking. The questionnaires measure usability through SUS~\cite{brooke1996sus}, presence through IPQ~\cite{schubert2001experience}, cognitive load through NASA-TLX~\cite{hart1988development}, and cybersickness through CSQ-VR~\cite{kourtesis2023cybersickness}. The eye-tracking analysis provided insights into engagement, cognitive behavior, and visual attention\revised{~\cite{holmqvist2011eye,zagermann2016measuring}}. Furthermore, we trained machine learning models to classify the locomotion technique based on eye-tracking features and explored how each feature contributed to the models’ decisions by applying explainable artificial intelligence using SHAP analysis~\cite{lundberg2017unified}. This process helped us identify the most influential gaze features in different locomotion techniques. 

Our findings confirm that controller-based teleportation is the fastest locomotion technique, as expected. Between the two voice-based techniques, we found no evidence of a difference in completion time under our task; however, the LLM-driven condition showed lower times in the second half, which may reflect potential learning effects over time. Across standardized self-report measures (SUS, IPQ, CSQ-VR, NASA-TLX), we similarly found no evidence of differences between conditions, suggesting that LLM-driven voice locomotion is feasible and provides user experience comparable to established baselines in this setting. In the eye-tracking analysis, we observed a descriptive trend toward longer fixation durations and shorter saccade durations in the LLM-driven teleportation condition, suggesting different visual behavior that may reflect more focused visual engagement during this task. Finally, we report an gaze-based classification analysis and SHAP feature attributions to highlight which eye-tracking features appear sensitive to locomotion mode.
Therefore, rather than demonstrating superiority, this paper contributes an initial design and feasibility evaluation of context-aware, LLM-driven hands-free locomotion for VR. To support reproducibility and enable future research, we provide our implementation and a supplementary video demonstrating the three locomotion techniques at \revised{\url{https://gitlab.lrz.de/hctl/vr-llm-locomotion}}.

\section{Related Work}
This section provides previous works in three folds, including locomotion techniques, hands-free interaction, and LLMs in VR.  

\subsection{Locomotion in VR}
Locomotion is one of the main components of user interaction in virtual environments, as it enables users to navigate and engage with digital spaces beyond their immediate physical boundaries. Without an effective locomotion mechanism, users are limited to static or highly constrained experiences, which may significantly diminish the immersive potential of virtual reality. Previous research proposed a broad spectrum of locomotion techniques in the literature~\cite{anderton2025teleportation,martinez2022research}, and they are generally categorized into five types: walking-based, steering-based, selection-based, manipulation-based, and automated locomotion. Walking-based approaches simulate natural gait and include methods such as redirected walking~\cite{bruder2012redirecting,langbehn2018evaluation}, omnidirectional treadmills~\cite{wehden2021slippery}, walking-in-place~\cite{tan2022understanding,feasel2008llcm}, and arm swinging~\cite{wilson2016vr}. Steering-based techniques allow users to control their movement direction through inputs such as joysticks~\cite{sargunam2018evaluating,coomer2018evaluating}, gaze or head orientation~\cite{qian2018look,bowman1997travel}, or voice commands~\cite{hombeck2023tell}, facilitating continuous movement through virtual spaces. Selection-based locomotion enables users to select a destination and instantly or gradually transport them to the destination. Among these, teleportation~\cite{bozgeyikli2016point} is one of the most widely adopted techniques, typically involving instant transportation, while others support smooth transitions toward selected points~\cite{liu2018increasing,medeiros2016effects}. Furthermore, manipulation-based methods allow users to directly control their virtual position by interacting with the scene or camera system~\cite{bozgeyikli2019locomotion,cho2017multi,raees2019ven}. In contrast, automated locomotion involves moving users along predefined routes or sequences under system control, often with minimal user input~\cite{paulo2020improving,rahimi2018scene}. 

In the context of voice-based teleportation-like locomotion for hands-free navigation, Ferracani et al.~\cite{ferracani2017natural} proposed a system for immersive museum experiences that enabled users, particularly those with motor impairments, to navigate by issuing semantic voice commands such as “I would like to see The Starry Night.” A rule-based grammar and ontological reasoning system interpreted these commands; however, the system required a manual definition of both object metadata and grammar rules, limiting its flexibility and scalability in more dynamic environments. Similarly, Calandra et al.~\cite{calandra2022comparison} evaluated speech-based navigation methods for training scenarios, including voice-only, voice with gaze, and hybrid approaches. While effective, their techniques relied on predefined destination names or required disambiguation logic through user interface panels and structured grammar templates, which may reduce naturalness and increase user cognitive load. Recent work has also evaluated hands-free teleportation alternatives that combine head or gaze pointing with non-controller confirmations such as dwell, mouth gestures, eye-winks, hand gestures, or voice~\cite{prithul2022evaluation,narbayev2025exploring}. These multimodal approaches provide important baselines for future comparisons, especially when the primary goal is hands-free teleportation rather than free-form natural language navigation.

\subsection{Hands-Free Interaction in VR}
Hands-free interaction is an essential feature of immersive VR systems~\cite{9995155}, especially when users must keep their hands free for primary tasks such as surgery or equipment operation~\cite{gallo2008toward}. Early multimodal conversational interfaces already showed that speech can be combined with pointing or gesture to support more natural spatial references in graphical interfaces~\cite{bolt1980put}. Monteiro et al.\cite{monteiro2021hands} conducted a comprehensive systematic review of hands-free interaction techniques in immersive VR, categorizing modalities based on input sources such as voice, eye gaze, and head movement. Among these, voice remains the most extensively studied modality due to its integration into commercial head-mounted displays (HMDs) and advancements in voice processing technologies~\cite{farinazzo2016usability, ferracani2017natural}.

Different voice-based interaction techniques have been proposed, including command-based and natural language processing (NLP)-based techniques. Command-based techniques operate on a limited set of predefined phrases (e.g., “go forward,” “stop”), which leads to higher command recognition accuracy and allows for offline or lightweight computational processing~\cite{schroeder2017presence}. These techniques are efficient and easy to learn, as users only need to memorize a small set of commands. However, their rigidity can be limiting, and they are not well-suited for complex tasks or dynamic interfaces, as they can be perceived as unintuitive by users. In contrast, NLP-based techniques enable users to issue free-form, natural language commands (e.g., “rotate by 45 degrees” or “select the purple circle”), offering more intuitive and expressive interaction~\cite{hepperle20192d}. These techniques are well-suited for symbolic input, parameterized commands, and narrative-driven tasks. However, they often depend on rule-based mechanisms that are sensitive to recognition errors caused by accent variation, phrasing differences, or background noise. Other voice-based approaches include symbolic input through voice-to-text~\cite{alfaro2018scientific} and non-verbal vocal cues~\cite{sra2018breathvr}, expanding the range of interaction possibilities.

Beyond voice, researchers have also utilized other hands-free approaches, including eye gaze for target selection or interaction with user interfaces~\cite{mardanbegi2019eyemrtk,blattgerste2018advantages}, often in combination with other modalities to mitigate issues such as the “Midas Touch” effect~\cite{jacob1991use}. Head gaze, facilitated by HMD tracking, is another intuitive input method that supports pointing and teleportation~\cite{prithul2022evaluation,narbayev2025exploring}. More novel modalities, such as facial expressions~\cite{ciftci2017partially}, body postures~\cite{gelsomini2020embodied}, and brain-computer interfaces (BCIs)~\cite{ma2018combining}, have gained attention for their potential to enhance accessibility and personalization. However, these approaches often require special hardware or user training, which can limit their widespread adoption.

\subsection{LLMs for VR}
The versatile computational capabilities of foundation models and LLMs make them suitable for handling complex tasks across different domains, such as medicine~\cite{Ayhan_etal_2025} and education~\cite{hou_etal_2024}. Considering the fact that LLMs can personalize user experiences often with few prompts and do not require extensive manual labor, such as in the form of annotations, similar to many domains, the VR community has also started to integrate those models into their workflows. In fact, integration of LLMs can facilitate inclusion and equity in virtual spaces and help increase users' engagement as they can provide infrastructure for intuitive interaction experience~\cite{bozkir2024embedding}. To this end, Buldu et al.~\cite{buldu2025cuify} focused on delivering a speech-based interaction framework that relies on LLMs in VR by combining different speech-to-text, text-to-speech, and language models. 
Skyba et al.~\cite{skyba2024towards} further investigated LLM-based natural language understanding for XR interaction, including navigation and spatial commands, showing both promise and challenges in interpreting complex spatial concepts.
Furthermore, Lau et al.~\cite{lau2024wrappedanansiswebunweaving} argued for using LLMs for personalization in VR spaces and found that personalization through LLMs in VR boosts engagement and learning interest, indicating the potential benefits of LLMs to support users. Furthermore, De La Torre et al.~\cite{de2024llmr} utilized LLMs to create interactive and virtual spaces, especially for producing and editing objects and scenes. The authors showed the effectiveness of their framework through a usability study in which participants provided positive feedback. This wide range of use cases of LLMs in VR shows their potential to make VR more engaging and accessible for users. In the context of this work, when hands-free locomotion is considered, one of the most straightforward ways to obtain the user's input is through voice input~\cite{hombeck2023tell}. However, previous research often considers fixed voice-based commands, limiting the user inputs' generalizability. We address this issue with LLMs by incorporating contextual information from the scene to facilitate hands-free and context-aware locomotion in VR. 

\section{Methodology}
In this study, we investigated the effectiveness of three distinct locomotion techniques in VR: teleportation, voice-based steering through fixed commands~\cite{hombeck2023tell}, and our proposed LLM-driven locomotion approach. To evaluate these techniques, we designed a virtual environment where participants were given a task that required them to navigate through the scene in our between-subjects design user study. 
We collected both objective and subjective measures. Objective data included eye-tracking metrics to assess attention and cognitive processes, while subjective data were gathered using the System Usability Scale (SUS), the Igroup Presence Questionnaire (IPQ), the Cybersickness in Virtual Reality Questionnaire (CSQ-VR), and the NASA Task Load Index (NASA-TLX).
These questionnaires measured usability, sense of presence, cybersickness, and cognitive load. In the following subsections, we describe the locomotion techniques and explain the details of our user study. \revised{The Ethics Commission of the Technical University of Munich} approved our study protocols and data collection procedures. 

\subsection{Locomotion Techniques}
In this study, we implemented three different locomotion techniques: controller-based teleportation, voice-based steering through fixed commands, and the proposed LLM-driven locomotion. Teleportation serves as a widely used baseline in VR locomotion studies and aligns with the instant transportation nature of the LLM-driven method. We included voice-based steering as a hands-free continuous-motion baseline using the same speech input modality as the LLM-driven method, while the LLM-driven method serves as a hands-free alternative that supports free-form, context-aware language. These conditions compare practical locomotion techniques rather than forming a full input-modality by movement-type factorial design. We therefore did not include controller-based steering, as it would mainly add a controller-driven continuous-motion condition outside the hands-free focus of this feasibility study.


\subsubsection{Controller-based Teleportation}
We use controller-based teleportation as the baseline locomotion method, representing a conventional and widely adopted technique in VR~\cite{bozgeyikli2016point,anderton2025teleportation}. This method is typically perceived as easy to learn and use, and we chose it because of its easy adaptability to most virtual environments. Users are generally comfortable with this locomotion technique~\cite{prithul2021teleportation}. In this method, the user points to a location using the controller button and is instantly teleported to the selected position. In the default implementation, a curved line is displayed to guide the user. A green arc indicates a valid teleportation target, while a red arc signals that teleportation to the selected area is not possible. When a valid location is selected, the user is immediately teleported to this location after participants pull their fingers from the button.\revised{ A single teleport could reach any valid ground within the user's line of sight, up to the ray interactor's default maximum reach of 30\,m.} This method minimizes motion-induced discomfort and is generally well-tolerated by users~\cite{bozgeyikli2016point}. 

\subsubsection{Voice-based Steering through Fixed Commands}
We utilize a voice-based steering locomotion method based on fixed commands, which is one of the most preferred hands-free techniques by users~\cite{hombeck2023tell}. It enables continuous directional control using a predefined set of voice commands. We selected this technique mainly because of its generic commands, which provide high adaptability across diverse virtual environments without requiring extensive customization. In contrast, other voice-based locomotion methods often rely on environment-specific configurations or require extensive manual annotations, which limits their scalability in large or dynamic virtual environments. For instance, traditional landmark-based approaches require semantic mapping and labeling for all recognizable objects in the scene, demanding significant manual effort. Similarly, number-grid-based systems~\cite{hombeck2023tell} become increasingly impractical as the environment grows in size, as they would require an overwhelming number of grid identifiers to cover the entire space.

In this technique, we utilized the VOSK speech recognition engine~\cite{shmyrev2020vosk} due to its robustness in handling a variety of voice commands and its demonstrated reliability in previous applications. Participants controlled movement through a predefined set of voice commands, including \texttt{<go forward>}, \texttt{<go back>}, \texttt{<turn left>}, \texttt{<turn right>}, and \texttt{<stop>}. Additional commands such as \texttt{<faster>} and \texttt{<slower>} are also supported, allowing users to adjust their walking speed dynamically across four discrete levels: 1.4\,m/s, 2.8\,m/s, 4.2\,m/s, and 5.6\,m/s. 
The baseline value of 1.4\,m/s approximates natural walking speed~\cite{OculusBestPractices} and was selected as the default comfort-oriented speed. Higher speeds were included to support more efficient traversal when participants chose to speed up, consistent with prior VR locomotion work using increased travel speeds~\cite{christou2017steering,caggianese2020freehand}. Prior immersive virtual travel work has also evaluated four discrete manual speed settings ranging from walking speed (1.5\,m/s) to fast running (8\,m/s), with intermediate values of 3.67\,m/s and 5.83\,m/s~\cite{freitag2016automatic}. Based on this precedent, our higher levels were included as optional speed-up settings for efficient traversal in the relatively large virtual town, rather than as assumed comfort-optimal speeds. We used integer multiples of the walking-speed baseline to make the \texttt{<faster>} and \texttt{<slower>} commands predictable and to avoid arbitrary speed increments.

Although this technique was designed as a hands-free alternative, participants used the same trigger button to start and stop recording in both voice conditions. This minimized voice-activation errors and kept activation consistent across the voice interfaces. Once a valid command such as \texttt{<go forward>} is provided, the user starts moving continuously at a constant speed. While directional changes could be controlled using the \texttt{<turn left>} and \texttt{<turn right>} commands, users are also free to rotate naturally using their physical head and body movements. Nevertheless, unlike teleportation, which involves instantaneous movement and minimizes sensory conflict, voice-based steering relies on continuous movement. This type of movement can create a mismatch between visual and vestibular signals, increasing the likelihood of VR-induced motion sickness~\cite{hombeck2023tell, mayor2019comparative}.

\subsubsection{LLM-driven Locomotion}
Our proposed LLM-driven locomotion technique introduces an approach to hands-free, natural language-based teleportation, enabling intuitive navigation in virtual environments through free-form voice commands. In contrast to conventional voice-based locomotion techniques~\cite{hombeck2023tell,calandra2022comparison,ferracani2017natural}, which rely on rigid command sets or rule-based grammar structures, our approach leverages LLMs to interpret open-ended and flexible instructions through contextual understanding. This approach eliminates the need for users to memorize specific commands or syntax, resulting in a more natural and engaging interaction experience. Moreover, the method offers inherent multilingual support without requiring separate command definitions for each language, thereby improving both accessibility for users and scalability for developers. To navigate the environment, users can employ naturally phrased voice commands, such as \texttt{<go to the red house>} or \texttt{<move 50 meters forward>}, which the system processes contextually to determine the intended destination. Thus, our LLM-driven approach can be considered a dynamically context-aware extension of landmark-based teleportation~\cite{hombeck2023tell,ferracani2017natural,calandra2022comparison}, as it eliminates the need for predefined object mappings by leveraging real-time scene understanding from runtime-extracted scene information, such as object names and tags.

To support this, we utilized the pipeline illustrated in Figure~\ref{fig:teaser}, which takes the system prompt, the user's voice input, and contextual scene information to construct a comprehensive user prompt. The system prompt primarily describes the purpose of the agent as a “navigation assistant in Unity” and explains the structure of the user prompts, specifying the rules and expected output format. It also provides three examples. The user prompt includes the user's command and contains explicit environmental context, including the positions of nearby landmarks, buildings, and vehicles currently visible to the user. This contextual information is obtained directly from objects that are both visible and have a specific tag, with their names extracted using a custom script in Unity. The LLM processes this information to generate precise target coordinates. Subsequently, these coordinates are dynamically mapped to the closest valid navigable location within the virtual environment based on predefined walkable areas. When generated coordinates correspond to multiple locations, such as intersections or corners, the system selects the street orientation that makes the smallest angle with the user’s head orientation to best match the intended movement trajectory. Additionally, to ensure robustness, the system incorporates several fallback mechanisms. If the LLM returns excessive text along with the location rather than only the expected coordinate, we dynamically extract the coordinate from the response. If the LLM fails to generate meaningful or interpretable output, or if the user's command is ambiguous, the user's position remains unchanged, effectively preventing unintended movement. Successful commands are visually indicated to the user by a curved line marking the calculated destination point, and after a two-second interval, the user is teleported to the new location. Similar to the voice-based steering through fixed commands, we use a trigger button to start and stop voice recording during the experiment to minimize the effect of design preferences. More details can be found in \revised{Appendices~\ref{sec:prompt}, \ref{sec:scene}, and \ref{sec:exec}}.

\begin{figure}[ht]
    \centering
    \includegraphics[width=0.5\columnwidth]{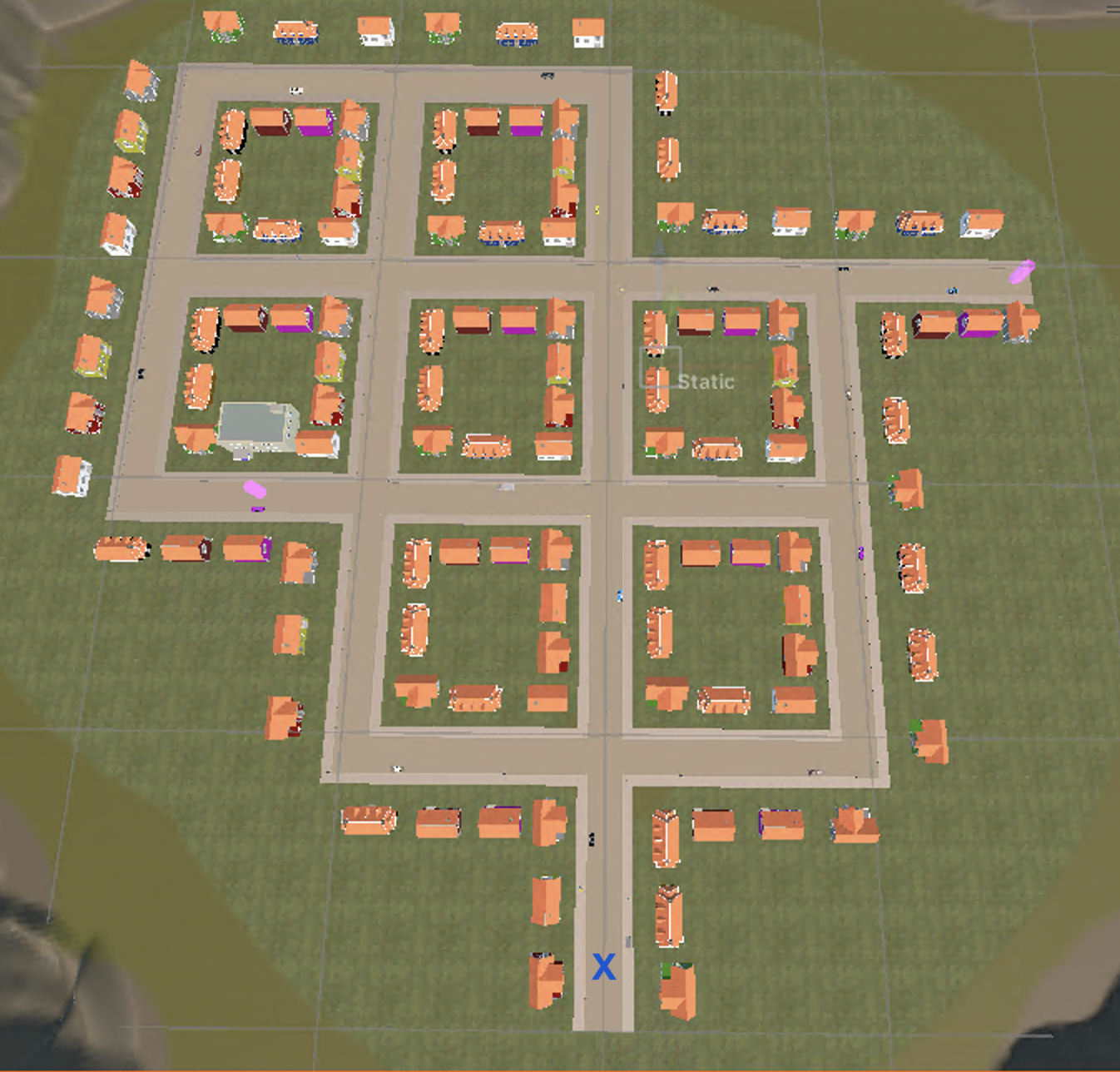}
    \caption{Top-down view of the virtual environment with the blue ‘X’ marking the start, and pink holograms indicate target locations. The pink hologram on the left corresponds to the first target, whereas the one on the right is the final target location.}
    \Description{Top-down view of the virtual neighborhood environment used in the navigation task. The road network forms a grid of residential blocks surrounded by houses and open areas. The blue “X” marks the participant’s starting position, while pink holographic markers indicate navigation targets. The left pink hologram corresponds to the first intermediate target, and the right pink hologram denotes the final target location within the environment.}
    \label{fig:env1}
    
\end{figure}

\subsection{User Study}
\subsubsection{Participants}
The study included data from 63 participants; however, due to technical issues, three participants were excluded, resulting in a final sample of 60 participants (20 per locomotion condition). All participants had normal or corrected-to-normal vision, with correction permitted only via contact lenses and all of them are compensated. The mean age is 27.1 years (SD = 5.33). The gender distribution is 60\% women (\(n = 36\)), 36.67\% men (\(n = 22\)), 1.67\% non-binary (\(n = 1\)), and 1.67\% who preferred not to disclose (\(n = 1\)). In terms of educational background, 46.67\% hold a bachelor’s degree or equivalent (\(n = 28\)), 33.33\% had a master’s degree or equivalent (\(n = 20\)), 13.33\% had completed high school or equivalent (\(n = 8\)), and 6.67\% hold a doctorate or equivalent (\(n = 4\)). Most participants were students (73.33\%, \(n = 44\)), while others were employed (20.00\%, \(n = 12\)), unemployed (5.00\%, \(n = 3\)), or categorized as other (1.67\%, \(n = 1\)). 98.33\% (\(n = 59\)) of participants had used LLMs before. Regarding VR experience, 70.00\% (\(n = 42\)) had used VR previously, although only 8.33\% (\(n = 5\)) own a VR device. 

\subsubsection{Apparatus}
We designed the environment in Unity and implemented three different locomotion techniques: teleportation using the VIVE controller, fixed command-based locomotion~\cite{hombeck2023tell} using the VOSK speech recognition engine~\cite{shmyrev2020vosk}, and LLM-driven locomotion utilizing the open-source CUIfy package~\cite{buldu2025cuify}, which provides an optimized pipeline for LLM-based interaction through speech-to-text models. For speech recognition, we used a locally hosted medium Whisper model~\cite{radford2023robust}, as it offers greater robustness compared to smaller versions, even though this introduced higher latency. For language understanding, we employed ChatGPT-4o~\cite{gpt4o} as the core LLM. We ran the experiments on a Varjo XR-3~\cite{varjo_xr3} HMD, connected to a desktop system equipped with an Intel Core i7-13700K processor, 32 GB of RAM, and an NVIDIA GeForce RTX 4080 GPU. We collected eye-tracking data with the XR-3’s built-in eye tracker at a 200 Hz sampling rate.

\subsubsection{Experimental Design}
We employed a between-subjects design, where each participant experienced only one of the three locomotion methods: teleportation, fixed voice command-based locomotion, or LLM-driven locomotion. This approach ensured that individual learning effects, fatigue, or cybersickness did not influence performance across different conditions. All participants navigated the same virtual environment. We conducted the study in a VR environment developed using Unity, designed to resemble a virtual small town with a simple street layout consisting of four vertical and four horizontal roads\revised{, spanning approximately $500 \times 500$\,m}, as shown in Figure~\ref{fig:env1}. We designed a two-step navigation task to evaluate each locomotion method within a task-oriented context. We first instructed participants to locate a bank in the virtual town. Each participant started at the same position, marked by the blue “X” in Figure~\ref{fig:env1}, and completed the task by finding the target locations at identical coordinates. Since the main goal was to compare navigation methods rather than participants’ ability to find targets, we made the task intuitive by placing a prominent pink hologram in front of the bank, making it easily identifiable.\revised{ The hologram was large enough to be seen from a distance as participants approached the relevant area, but it was not visible from every location in the town.} Upon reaching the bank\revised{, detected automatically when the participant moved within a proximity radius of 10\,m of the target}, the first hologram disappeared, and a second one appeared at the town’s exit as a second target. Participants were then required to find and reach this exit point to complete the navigation task. In the LLM-based mode, participants could travel visible distances of \revised{up to} 50 meters\revised{ per command, as both the navigable street segments and the surrounding objects collected as scene context were limited to this same range around the user}. Therefore, they could not directly say ‘go to the bank.’.

\subsubsection{Procedure}
We first welcomed the participants and asked them to provide written consent to participate in the experiment. Next, we explained the (randomly) assigned locomotion condition to them in detail. Participants then wore the HMD, and the session began with a demo scene designed to help them become familiar with the locomotion technique and VR. The demo scene used the same virtual environment, starting from the same initial position, and included a single target located at the first corner, which was easily visible and close to the starting point. Eye-tracking calibration was also performed during this phase. Once participants felt comfortable about the VR and controls, following the demo, the main experiment commenced. As instructed, participants were supposed to navigate the virtual environment and reach two holograms placed at specific locations. After completing the task, participants completed questionnaires, including the NASA-TLX, SUS, IPQ, CSQ-VR, and a demographic survey. Later, they received €7.5 for the 30-minute experimental session.

\subsubsection{Measurements}
We conducted a comprehensive analysis of each locomotion technique using statistical methods. We logged task completion time for the intermediate and final targets. We collected eye-tracking data to objectively evaluate participants’ engagement and cognitive behavior.  In addition, we collected their feedback through standardized questionnaires, including the SUS, IPQ, CSQ-VR, and NASA-TLX. We describe these details in the following.

\paragraph{Eye Tracking Analysis}
To evaluate the locomotion techniques, we conducted an eye-tracking analysis, as eye tracking has been widely used to assess user engagement, attention, behavior, and cognitive processes during interaction~\cite{holmqvist2011eye,zagermann2016measuring}. We classified samples as either fixations or saccades from the raw eye-tracking data. Fixations represent moments when the eyes remain steadily focused on a specific point, while saccades are rapid eye movements between fixation points. We analyzed several features to understand users’ visual attention and cognitive states. The fixation rate, which reflects the frequency of attentional shifts, is commonly used to assess how users allocate their attention during a task~\cite{eckstein2017beyond}. Mean fixation duration, which represents the length of time users dwell on specific points, has been linked to deeper cognitive processing and enhanced information encoding~\cite{just1980theory}. The mean saccade duration captures the speed of gaze transitions, indicating the efficiency of visual search behavior\revised{~\cite{holmqvist2011eye,eckstein2017beyond}}. Pupil diameter provides insights into the mental effort and arousal, correlating with the cognitive workload, fatigue, and engagement~\cite{beatty1982task,hopstaken2015window}. Additionally, we identified blinks and extracted both blink frequency and duration, as both are associated with cognitive load and visual fatigue~\cite{benedetto2011driver}. \revised{Blink frequency and duration were included in the feature set used for the classification analysis (Table~\ref{tbl_features}).}

\paragraph{System Usability Scale (SUS)}
To assess usability, we employed the System Usability Scale (SUS)~\cite{brooke1996sus}, one of the most widely used tools for evaluating system usability. The questionnaire consists of 10 items, with participants rating each statement on a 5-point Likert scale, ranging from 1 (strongly disagree) to 5 (strongly agree). The SUS score is calculated by summing the scores for each question and then scaling the total, resulting in a final score ranging from 0 to 100, where higher scores indicate better usability.

\paragraph{Igroup Presence Questionnaire (IPQ)}
To assess the sense of presence in the virtual environment and the influence of each locomotion technique on it, we used the Igroup Presence Questionnaire (IPQ)~\cite{igroup_ipq}. While based on the theoretical framework introduced by Schubert et al.~\cite{schubert2001experience}, the IPQ consists of 14 questions designed to evaluate presence. It measures four dimensions: Spatial Presence (SP), Involvement (INV), Experienced Realism (REAL), and General Presence (PRES). Each question is rated on a 7-point Likert scale, and the mean score is calculated for each category.

\paragraph{CyberSickness in Virtual Reality Questionnaire (CSQ-VR)}
To evaluate cybersickness, we used the Cybersickness in Virtual Reality Questionnaire (CSQ-VR)~\cite{kourtesis2023cybersickness}, which is designed to assess both the intensity and type of cybersickness symptoms, including nausea, disorientation, and oculomotor discomfort, experienced during VR exposure. The questionnaire comprises six items, each rated on a 7-point Likert scale.

\paragraph{NASA-Task Load Index (NASA-TLX)}
To assess the impact of the locomotion technique on participants’ cognitive load, we employed the NASA-TLX~\cite{hart1988development}, a widely used tool for evaluating perceived workload across multiple dimensions.

\begin{table}[h]
\small
\centering
\caption{Criteria for fixation and saccade detection.}
\revised{\Description{Table of event-detection thresholds. A sample is classified as a fixation when head velocity is below 7 degrees per second and gaze velocity below 30 degrees per second, with duration between 80 and 500 milliseconds. A sample is classified as a saccade when gaze velocity exceeds 40 degrees per second, with duration between 20 and 70 milliseconds.}}
\begin{tabular}{@{}p{0.1\columnwidth} p{0.15\columnwidth} p{0.2\columnwidth}@{}}
\toprule
\textbf{Event} & \textbf{Velocity ($v$)} & \textbf{Duration ($\Delta$)} \\
\midrule
Fixation & 
\begin{tabular}[t]{@{}l@{}} 
$v_{head} < 7^\circ/s$ \\ 
$v_{gaze} < 30^\circ/s$ 
\end{tabular} 
& 
\begin{tabular}[t]{@{}l@{}} 
$ \Delta_{fixation} > 80 \ ms $ \\ 
$\Delta_{fixation} < 500 \ ms$ 
\end{tabular} \\
\addlinespace
Saccade & 
$v_{gaze} > 40^\circ/s$ & 
\begin{tabular}[t]{@{}l@{}} 
$\Delta_{saccade}>20 \ ms $ \\ 
$\Delta_{saccade} < 70 \ ms$ 
\end{tabular} \\
\bottomrule
\end{tabular}
\label{table:criteria}
\end{table}

\subsubsection{Data Processing}
To analyze eye-tracking data and pupillometry data, we first need a multi-step data processing pipeline. First, based on gaze velocity, we classified eye movement data into fixations and saccades using the Velocity-Threshold Identification (I-VT) algorithm~\cite{salvucci2000identifying,kasneci2024introduction}. Stable gaze movements are classified as fixations, while faster movements are labeled as saccades. We also incorporated head movement data to improve the fixation detection reliability. We only considered a fixation when both the gaze and head remained stable, similar to previous works~\cite{gao2021digital,agtzidis2019360,gao2022eye}. We present the detailed criteria for classifying fixations and saccades by including velocity and duration thresholds in Table~\ref{table:criteria}.

To ensure an accurate interpretation of pupil diameter, we applied a Savitzky-Golay filter~\cite{savitzky1964smoothing} to smooth the raw signal and reduce short-term noise. Subsequently, we performed a divisive baseline correction using a one-second pre-stimulus interval~\cite{mathot2018safe}, allowing for normalization across participants and experimental trials. Pupil diameter was used as an indicator of objective cognitive load and engagement, similar to prior virtual reality studies~\cite{gao2021digital,bozkir2019person,bozkir2019assessment}.

Additionally, we analyzed blinks using the Varjo XR-3's eye openness data. We identified a blink when the eye openness value reached zero, following a consistent decrease in eye openness, indicating a natural eyelid closure. To minimize false positives caused by tracking loss, we discarded blink candidates that did not exhibit a decreasing eye openness trend before reaching zero.

\subsubsection{Analysis}
We conducted a series of statistical analyses on both questionnaire responses (e.g., SUS, NASA-TLX) and eye-tracking measures to examine differences across experimental conditions. For each dependent variable, we performed a one-way Analysis of Variance (ANOVA) to evaluate the effect of the locomotion technique (teleportation, LLM-driven locomotion, and voice-based steering). When the ANOVA indicated a statistically significant effect, we used Tukey’s Honest Significant Difference (HSD) post-hoc test to determine which pairs of conditions differed significantly, thereby controlling the family-wise error rate across all comparisons.

Before conducting ANOVA, we checked the assumptions of normality with the Shapiro–Wilk test and homogeneity of variance with Levene’s test. In cases where these assumptions were violated, we conducted the non-parametric Kruskal-Wallis H test as an alternative to ANOVA. We conducted all statistical tests using $\alpha = 0.05$. Where relevant, we also report the effect sizes (e.g., $\eta^2$) to support the interpretation of findings.

\subsubsection{Model Building}
We conducted detailed analyses of eye-tracking data to investigate how different locomotion techniques influence user behavior. These analyses aim to reveal potential cognitive and attentional differences among participants using different locomotion methods. Additionally, we trained classification models to predict the locomotion condition based on eye-tracking features. This exploratory analysis helps identify which features are most affected by the locomotion technique and which features are the most predictive or distinctive in differentiating between the locomotion techniques. The feature set was selected from commonly used gaze, blink, and pupil metrics associated with attention, visual search, fatigue, and cognitive load~\cite{holmqvist2011eye,just1980theory,beatty1982task,benedetto2011driver,bahill1975main}. Non-linear pupillary responses to cognitive load are well-established~\cite{beatty1982task,payne1968workload}. To enable the Logistic Regression L2 classifier to capture these quadratic relationships in pupil dynamics, we applied a degree-2 polynomial expansion (squared transforms) to the pupil diameter features. We provide the full list of features used in Table~\ref{tbl_features}.

\begin{table}[htbp]
\centering
\small
\setlength{\tabcolsep}{4pt}
\caption{Eye-tracking feature list.}
\label{tbl_features}
\revised{\Description{Table listing the eye-tracking features used for classification, grouped into fixation, saccade, blink, and pupil-diameter metrics, together with the statistical summaries (such as mean, standard deviation, minimum, maximum, sum, and Shannon entropy) computed for each feature.}}
\makebox[\linewidth][c]{%
\begin{tabular}{>{\raggedright\arraybackslash}p{3.9cm} >{\raggedright\arraybackslash}p{9.2cm}}
\toprule
\textbf{Features} & \textbf{Statistical Metrics} \\
\midrule
Number of fixations & Total number of fixations \\
Fixation duration & Mean, Std, Min, Max, Sum, and Shannon entropy of fixation durations \\
\midrule
Number of saccades & Total number of saccades \\
Saccade duration & Mean, Std, Min, Max, and Sum of saccade durations \\
Inter-saccadic interval & Mean and Std of inter-saccadic intervals  \\
Saccade peak velocity & Mean, Std, Log-Std, Min, Max of saccade peak velocities \\
Saccade amplitude & Mean, Std, Min, and Max of saccade amplitudes \\
Saccade direction & 8-bin angular histogram of saccade directions \\
Saccade main sequence slope & Slope of $\log_{10}$ peak velocity vs.\ $\log_{10}$ amplitude~\cite{bahill1975main} \\
\midrule
Sac-fix ratios & Pairwise sac--fix duration/count ratios \\
\midrule
Number of blinks & Total number of blinks \\
Blink duration & Mean, Std, Min, and Max of blink durations \\
\midrule
Pupil diameter & Mean, Std, Min, Max and second-order polynomial expansion of pupil statistics \\
\midrule
Event count & Total number of event count in a window \\
\bottomrule
\end{tabular}
}
\end{table}

To this end, we evaluated five classifiers: L2-regularized logistic regression, linear support vector machine (Linear SVM)~\cite{cortes1995support}, Ridge classifier with L2 regularization~\cite{hoerl1970ridge}, L1-regularized logistic regression, and Linear Discriminant Analysis (LDA). All classifiers were trained on features standardized with a RobustScaler~\cite{scikit-learn}, fitted on the training folds only to prevent test leakage. The classification was performed on features extracted from 3-second windows. We empirically evaluated all integer window sizes from 1 to 20 seconds and selected 3 seconds as the best compromise between temporal resolution and stable estimation of fixation, saccade, blink, and pupil statistics (see Appendix~\ref{app:model}). Also, shorter windows were not feasible because they frequently yielded zero events for lower-frequency categories such as blinks, making per-window aggregation unreliable. Window-level features were averaged across all windows of a participant to obtain a single feature vector per participant; the 3-second window also provided the most stable participant-level feature estimates. We used leave-one-participant-out cross-validation (LOOCV) across all $60$ participants; the split was performed at the participant level rather than the window level, so no information from the test participant was available during training, ensuring strict between-subject generalization. We selected hyperparameters by evaluating a small grid of configurations and reporting the best-performing one. To understand the model's decision-making processes, we applied SHAP (SHapley Additive exPlanations)~\cite{lundberg2017unified} to interpret feature importance in a transparent and model-agnostic manner. SHAP provides a detailed attribution of feature contributions toward classification outcomes by quantifying each feature's marginal impact on the model predictions, helping us identify which aspects of gaze behavior, such as fixation duration, saccade amplitude, or pupil diameter, are most influenced by the employed locomotion technique and may also support personalized VR experiences.

\section{Results}

\subsection{Performance Analysis} \label{sec:performance_analysis}
\begin{figure}[ht]
    \centering
    \includegraphics[width=0.31\columnwidth]{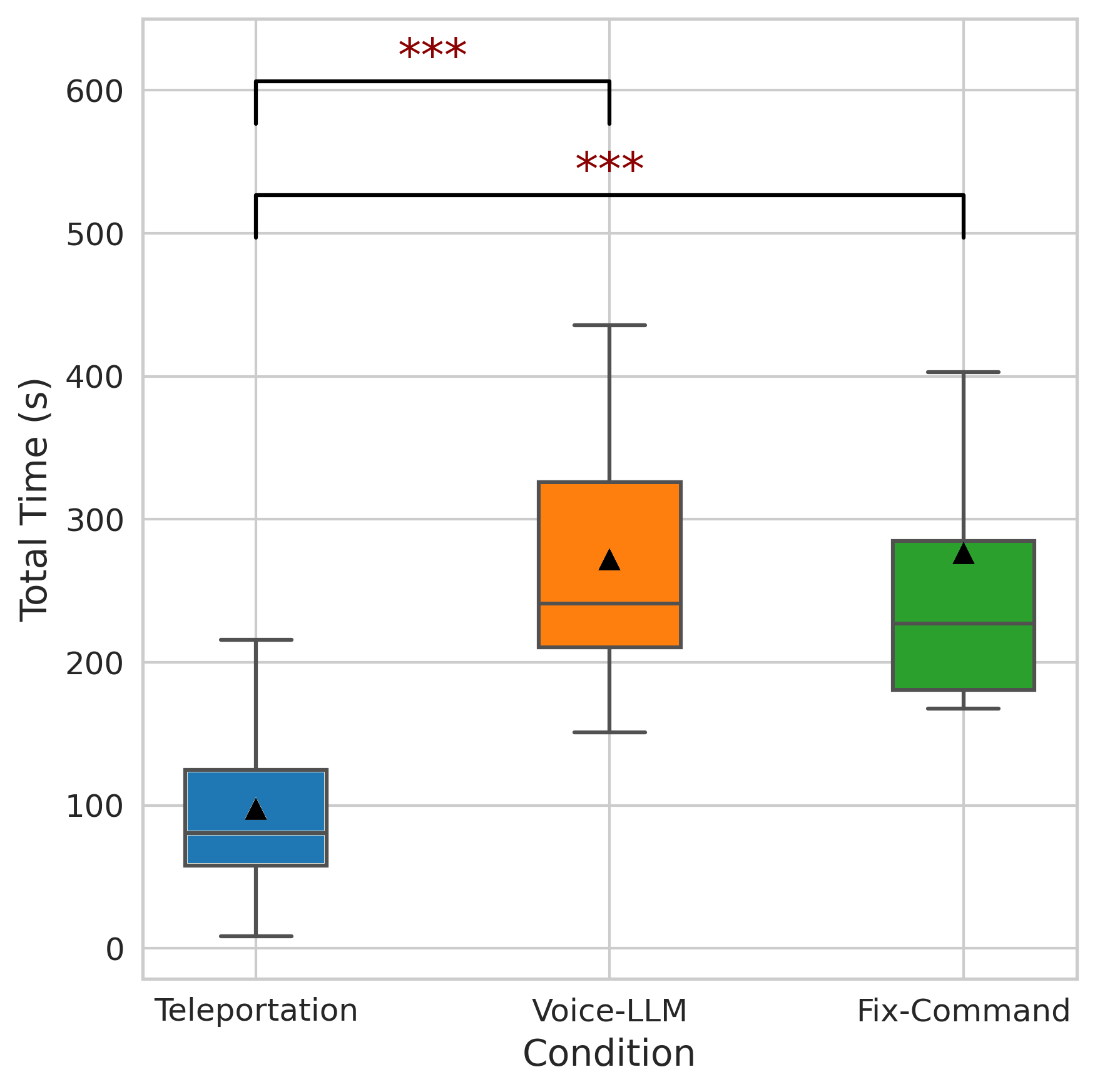}
    \caption{Results for task completion times.}
    \Description{Boxplots showing task completion times across three navigation conditions: Teleportation, Voice-LLM, and Fixed-Command. The y-axis denotes total completion time in seconds. Boxes represent the interquartile range, horizontal lines indicate medians, whiskers show the data range, and black triangles denote mean values. Teleportation yields the shortest completion times, while Voice-LLM and Fixed-Command result in longer durations. Statistically significant differences between conditions are indicated by brackets with asterisks (***).}
    \label{fig:time2end}
\end{figure}

Figure~\ref{fig:time2end} presents the participants’ task completion times across the three locomotion conditions. On average, participants completed the task in $M = 96.45$, $SD = 57.76\,\text{s}$ using teleportation, $M = 270.77$, $SD = 84.67\,\text{s}$ with the LLM-driven locomotion, and $M = 275.32$, $SD = 183.38\,\text{s}$ with the fixed voice command approach. As expected, teleportation was the fastest method, serving as the baseline for comparison. A one-way ANOVA revealed a significant effect of locomotion condition on task completion time, $F(2, 57) = 14.14$, $p < .001$, with a large effect size ($\eta^2 = .332$). Post-hoc tests showed teleportation was significantly faster ($p < .001$), while the LLM-driven and fixed command methods showed no significant difference ($p = .997$).

We further analyzed the time required to reach each of the two sequential targets to explore potential learning effects. For the first target (the bank in the virtual town), participants in the voice-based steering condition reached it in $M = 121.96$, $SD = 54.69\,\text{s}$, while those in the LLM-driven condition took $M = 137.61$, $SD = 77.26\,\text{s}$. From the bank to the second target, the LLM-driven condition completed it in $M = 133.16$, $SD = 63.64\,\text{s}$, whereas the voice-based steering condition required $M = 153.36$, $SD = 143.05\,\text{s}$. \revised{In terms of completion time, a tendency toward improvement from the first to the second segment was observed only in the LLM-driven condition, whereas the voice-based steering condition took descriptively longer in the second segment; in either case, no statistically significant differences were found.}

In addition to completion times, we also measured system latency for LLM-driven locomotion, as teleportation and voice-based steering were nearly instantaneous. The speech-to-text component, using a locally hosted medium-sized Whisper model, had an average processing time of $M = 0.48$, $SD = 0.06\,\text{s}$, while the LLM component, utilizing the ChatGPT-4o API, had an average response time of $M = 0.97$, $SD = 0.74\,\text{s}$. In total, the average end-to-end processing time remained under 1.5 seconds ($M = 1.44$, $SD = 0.74\,\text{s}$). Additionally, to characterize ASR behavior, we calculated word error rate (WER) and exact-match rates for the two voice-based conditions. The Whisper-based LLM pipeline produced a mean WER of $0.038$ with an exact-match rate of $88.24\%$, while the VOSK fixed-command pipeline produced a mean post-mapping WER of $0.066$ with an exact-match rate of $86.05\%$. For VOSK, this post-mapping score reflects the command actually sent to the locomotion system, because recognized speech was mapped to the closest valid command. These results should be interpreted with caution because VOSK was evaluated over a small fixed command vocabulary, whereas Whisper was used for free-form speech in the LLM-driven condition.

\subsection{Eye-Tracking Analysis}
\begin{figure*}[ht]
    \centering
    \begin{minipage}{0.23\textwidth}
        \centering
        \includegraphics[width=\linewidth]{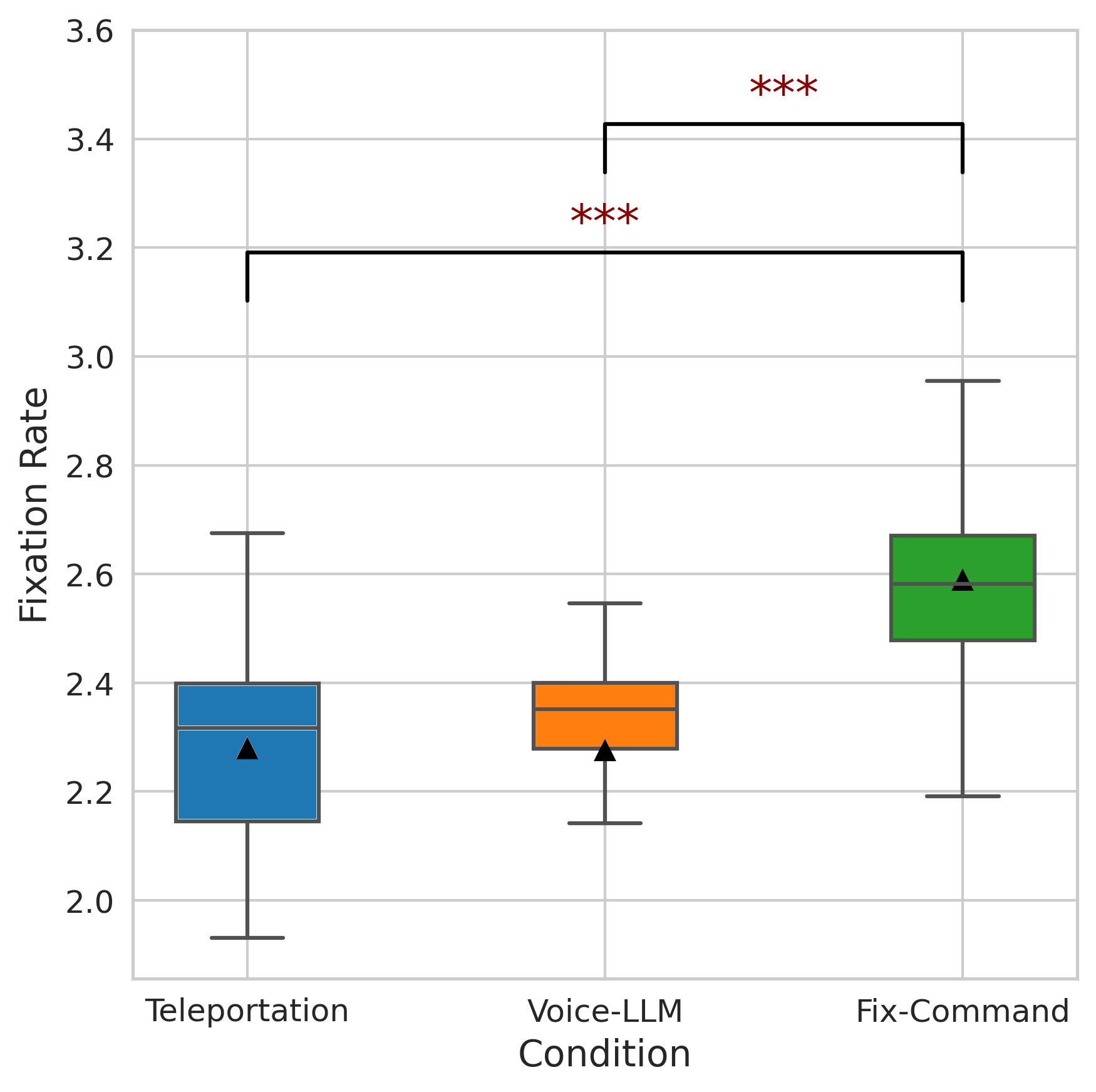}
        \caption{Results for fixation rates.}
        \Description{Boxplots show normalized fixation rates across the three navigation conditions. Fixation rates are lowest for Teleportation and increase for Voice-LLM and Fixed-Command. Statistically significant differences are indicated by brackets and asterisks (\*\*\*).}
        \label{fig:anova_boxplot_fixation_count_norm}
    \end{minipage}
    \hfill
    \begin{minipage}{0.23\textwidth}
        \centering
        \includegraphics[width=\linewidth]{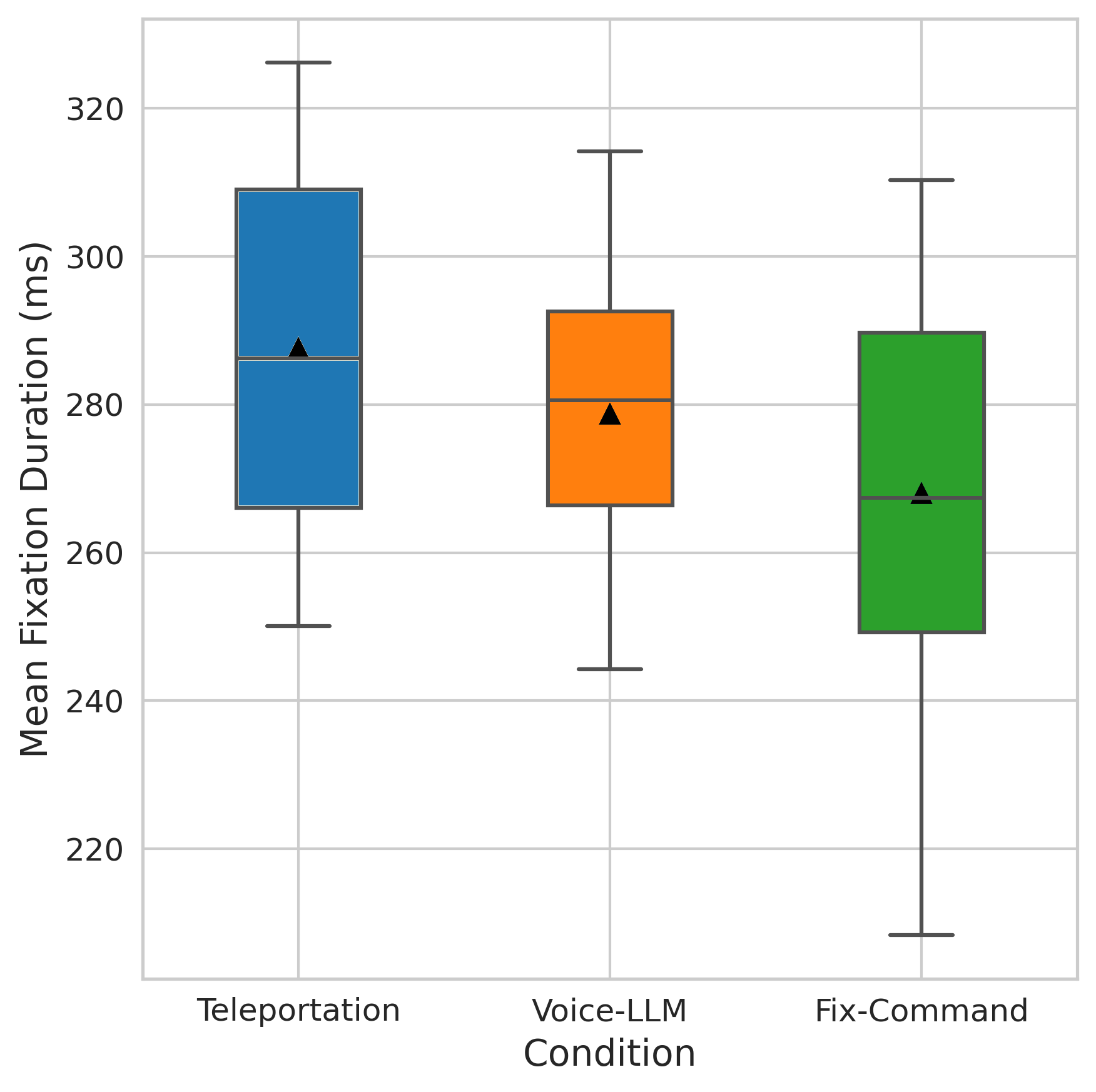}
        \caption{Results for mean fixation duration.}
        \Description{Boxplots illustrate mean fixation duration (ms) across navigation conditions. Teleportation shows longer fixations, while Voice-LLM and Fixed-Command exhibit shorter fixation durations.}
        \label{fig:mean_fixation_duration}
    \end{minipage}
    \hfill
    \begin{minipage}{0.23\textwidth}
        \centering
        \includegraphics[width=\linewidth]{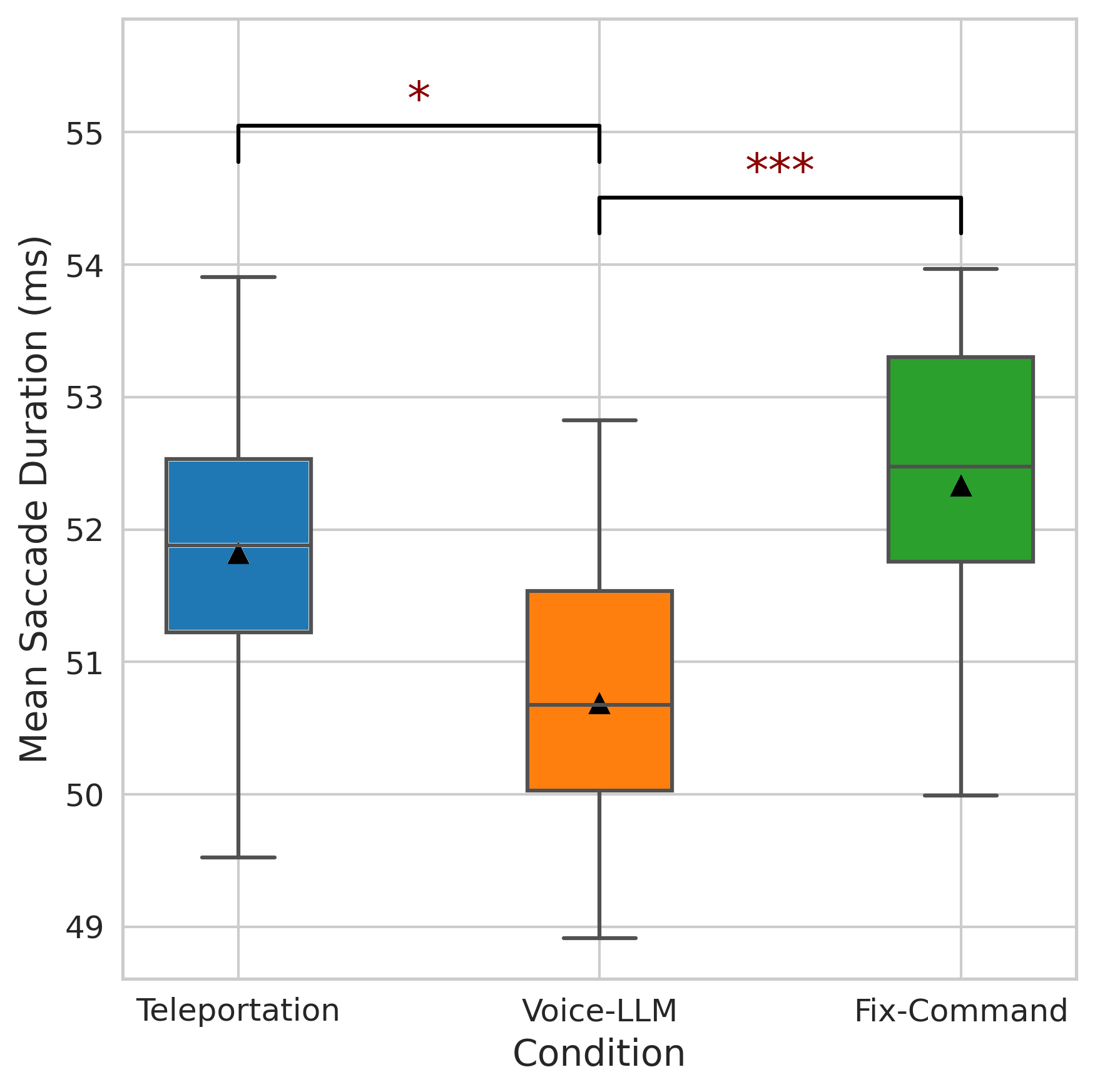}
        \caption{Results for mean saccade duration.}
        \Description{Boxplots present mean saccade duration (ms) for each condition. Voice-LLM produces shorter saccades compared to Teleportation and Fixed-Command, with significant differences indicated by asterisks.}
        \label{fig:mean_saccade_duration}
    \end{minipage}
    \hfill
    \begin{minipage}{0.23\textwidth}
        \centering
        \includegraphics[width=\linewidth]{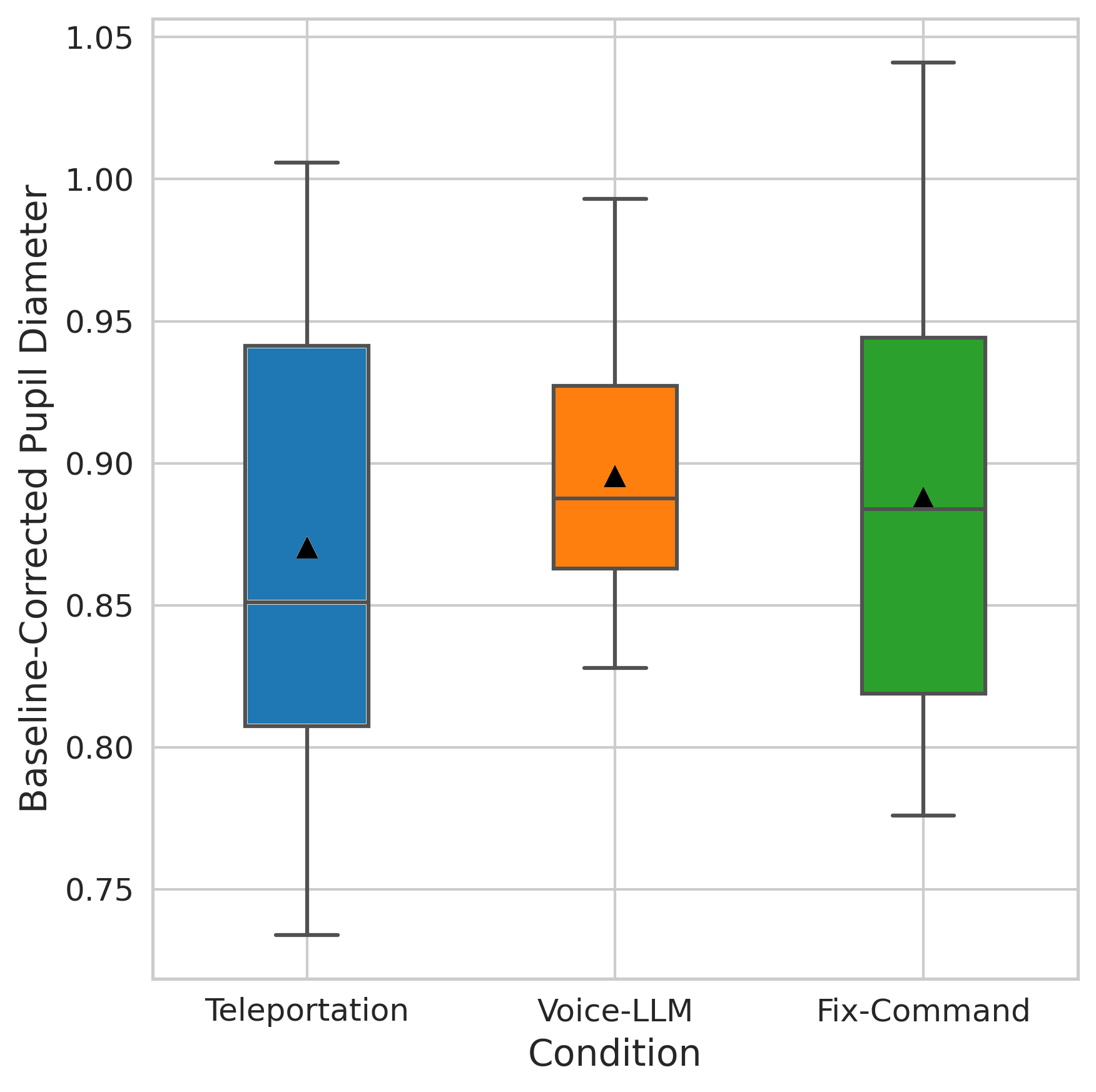}
        \caption{Results for mean pupil diameter.}
        \Description{Boxplots show baseline-corrected mean pupil diameter across conditions. No significant differences are observed between navigation techniques.}
        \label{fig:mean_pupil_diameter}
    \end{minipage}
\end{figure*}

\paragraph{Fixation rate}
To assess cognitive engagement under different locomotion techniques, we analyzed the fixation rate, defined as the number of fixations per second, as shown in Figure~\ref{fig:anova_boxplot_fixation_count_norm}. ANOVA revealed a statistically significant effect of locomotion type on fixation rate, $F(2, 57) = 11.77$, $p < .001$, with a large effect size ($\eta^2 = .292$), indicating that the type of locomotion significantly influenced visual attention patterns. Additionally, the statistics indicated that the voice-based steering condition resulted in the highest fixation rate ($M = 2.58$, $SD = 0.17$), followed by teleportation ($M = 2.28$, $SD = 0.22$), and voice LLM ($M = 2.27$, $SD = 0.28$). Post-hoc comparisons using Tukey’s Honest Significant Difference (HSD) test showed that both teleportation and voice LLM conditions had significantly lower fixation rates compared to the voice-based steering condition ($p < .001$ for both comparisons). However, no statistically significant difference was found between the teleportation and voice LLM conditions ($p = .99$).

\paragraph{Mean Fixation Duration}
Mean fixation durations are presented in Figure~\ref{fig:mean_fixation_duration}, and all values are reported in milliseconds. ANOVA revealed no statistically significant differences between conditions, $F(2, 57) = 2.86$, $p = .065$. Among the conditions, teleportation showed the highest mean fixation duration ($M = 287.42$, $SD = 25.67$ms), followed by LLM-driven locomotion ($M = 278.54$, $SD = 24.45$ms), and voice-based steering with the lowest mean duration ($M = 267.85$, $SD = 27.71$ms).
\paragraph{Mean Saccade Duration}
Mean saccade durations for each condition are presented in Figure~\ref{fig:mean_saccade_duration}, and all values are reported in milliseconds. ANOVA revealed a statistically significant effect of locomotion type on mean saccade duration, $F(2, 57) = 8.75$, $p < .001$, with a large effect size ($\eta^2 = .235$). Descriptive statistics showed the highest mean saccade duration in voice-based steering ($M = 52.33$, $SD = 1.28$ms), followed by teleportation ($M = 51.87$, $SD = 1.24$ms), and the lowest in the LLM-driven condition ($M = 50.76$, $SD = 1.49$ms). Post-hoc tests showed mean saccade duration was significantly lower in the LLM-driven condition than in teleportation ($p = .018$) and voice-based steering ($p < .001$), with no difference between the latter two ($p = .418$).

\paragraph{Pupil Diameter}
We analyzed the mean baseline-corrected pupil diameter across the three locomotion conditions to assess participants' cognitive load as shown in Figure~\ref{fig:mean_pupil_diameter}. ANOVA revealed no statistically significant differences between the conditions, $F(2, 57) = 2.38$, $p = .103$. Although not statistically significant, the descriptive statistics suggest a pattern. The teleportation condition caused the lowest mean pupil diameter ($M = 0.86, SD = 0.08$), while the voice-based conditions showed higher and relatively similar values (Voice LLM: $M = 0.90, SD = 0.06$; Voice-Based Steering: $M = 0.88, SD = 0.07$). This pattern may reflect slightly higher cognitive effort in the voice-based hands-free methods compared to teleportation. 

\subsection{Questionnaire Analysis}

\subsubsection{CSQ-VR Questionnaire}
\begin{figure}[ht]
    \centering
    \includegraphics[width=0.6\columnwidth]{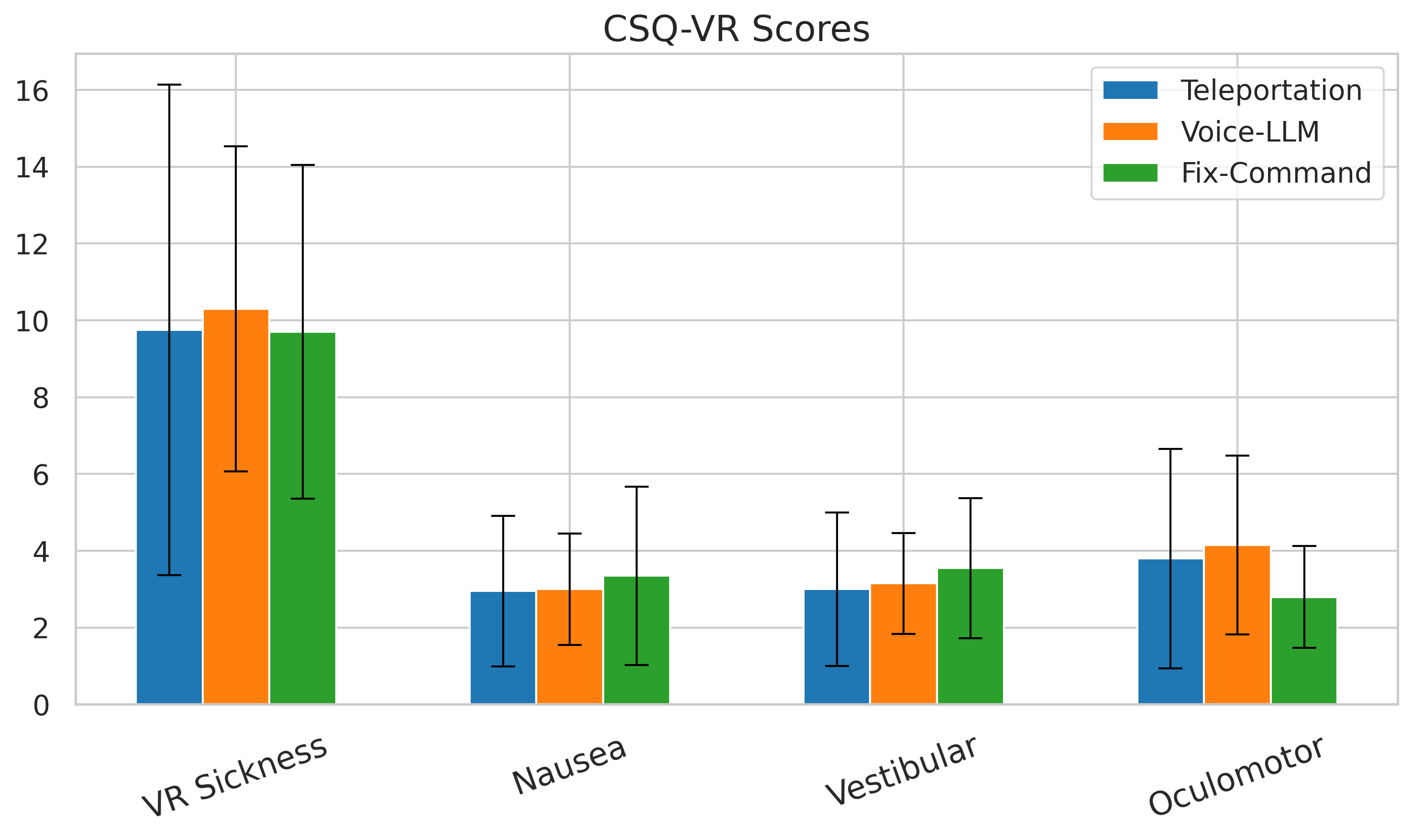}
    \caption{Results for \revised{CSQ-VR} Scores. \revised{Error bars denote $\pm$1 SD across participants.}}
    \Description{Bar plots showing mean CSQ-VR subscale scores for VR Sickness, Nausea, Vestibular, and Oculomotor symptoms across the three navigation conditions: Teleportation, Voice-LLM, and Fixed-Command. Error bars indicate variability across participants. Overall, scores are comparable across conditions, suggesting no substantial differences in cybersickness-related symptoms between navigation techniques.}
    \label{fig:csqvr_scores}
\end{figure}

Figure~\ref{fig:csqvr_scores} illustrates the CSQ-VR scores across locomotion techniques, covering overall cybersickness as well as the subcategories of nausea, vestibular, and oculomotor symptoms. The general score range for the CSQ-VR questionnaire is between 0 and 12 for nausea, vestibular, and oculomotor symptoms, and between 0 and 36 for overall sickness. Overall, participants reported low levels of cybersickness across all conditions, with no statistically significant differences observed. Teleportation showed the lowest scores in overall sickness ($M = 9.75$, $SD = 6.39$), nausea ($M = 2.95$, $SD = 1.96$), and vestibular discomfort ($M = 3.00$, $SD = 2.00$), suggesting a generally more comfortable experience. For overall sickness, the LLM-driven condition showed the highest scores ($M = 10.30$, $SD = 4.23$); however, the voice-based steering condition resulted in the highest levels of nausea ($M = 3.35$, $SD = 2.32$) and vestibular symptoms ($M = 3.55$, $SD = 1.82$), while the LLM-driven condition fell between the two. Interestingly, in the oculomotor discomfort, the voice-based steering condition showed the lowest scores ($M = 2.80$, $SD = 1.32$), whereas both teleportation ($M = 3.80$, $SD = 2.86$) and the LLM-driven ($M = 4.15$, $SD = 2.32$) reported higher values. No statistically significant differences were found between conditions.

\subsubsection{IPQ Scores}
\begin{figure}[ht]
    \centering
    \includegraphics[width=0.6\columnwidth]{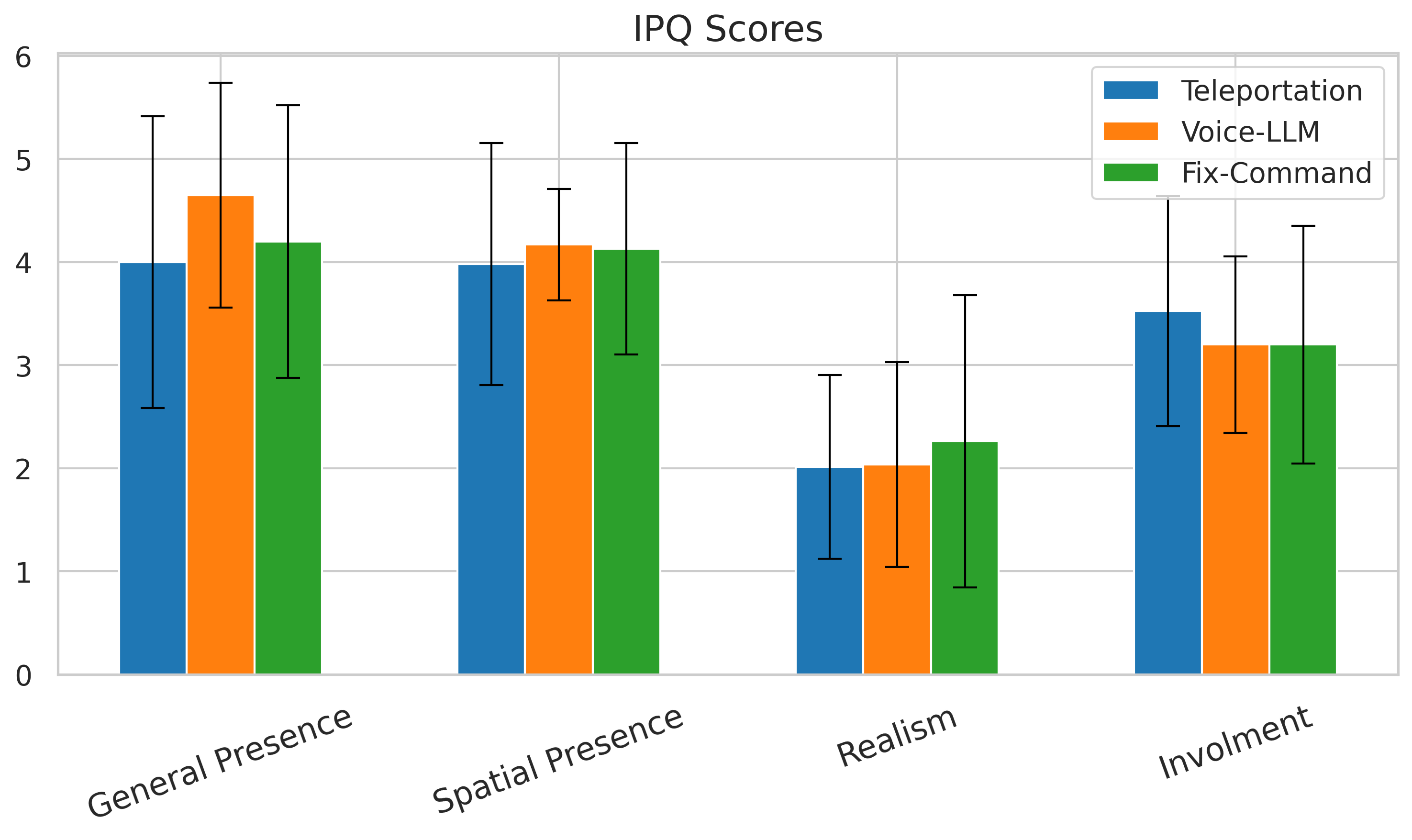}
    \caption{\revised{Results for IPQ Scores.}}
    \revised{\Description{Bar plots showing mean Igroup Presence Questionnaire (IPQ) scores across the three navigation conditions (Teleportation, Voice-LLM, and Fixed-Command) for General Presence, Spatial Presence, Realism, and Involvement. Error bars indicate variability across participants. The LLM-driven condition shows the highest general and spatial presence, voice-based steering the highest realism, and teleportation the highest involvement.}}
    \label{fig:ipq_scores}
\end{figure}

The IPQ results, shown in Figure~\ref{fig:ipq_scores} \revised{(error bars denote $\pm$1 SD across participants)}, present scores across four dimensions of presence: general presence, spatial presence, realism, and involvement, for each locomotion condition. Scores range from 0 to 6. The LLM-driven locomotion method received the highest ratings in both general presence ($M = 4.65$, $SD = 1.09$) and spatial presence ($M = 4.17$, $SD = 0.54$), indicating that participants experienced a stronger sense of immersion in the virtual environment. In the realism dimension, voice-based steering scored highest ($M = 2.26$, $SD = 1.42$), suggesting participants perceived the environment as more realistic with this method. For involvement, teleportation resulted in the highest score ($M = 3.53$, $SD = 1.11$), while the LLM-driven locomotion ($M = 3.20$, $SD = 0.86$) and voice-based steering ($M = 3.20$, $SD = 1.15$) showed similar values. However, we found no statistically significant differences across the conditions ($p = .268$).

\subsubsection{NASA-TLX Questionnaire}

\begin{figure}[ht]
    \centering
    \begin{minipage}[t]{0.45\columnwidth}
        \centering
        \includegraphics[width=0.6\linewidth]{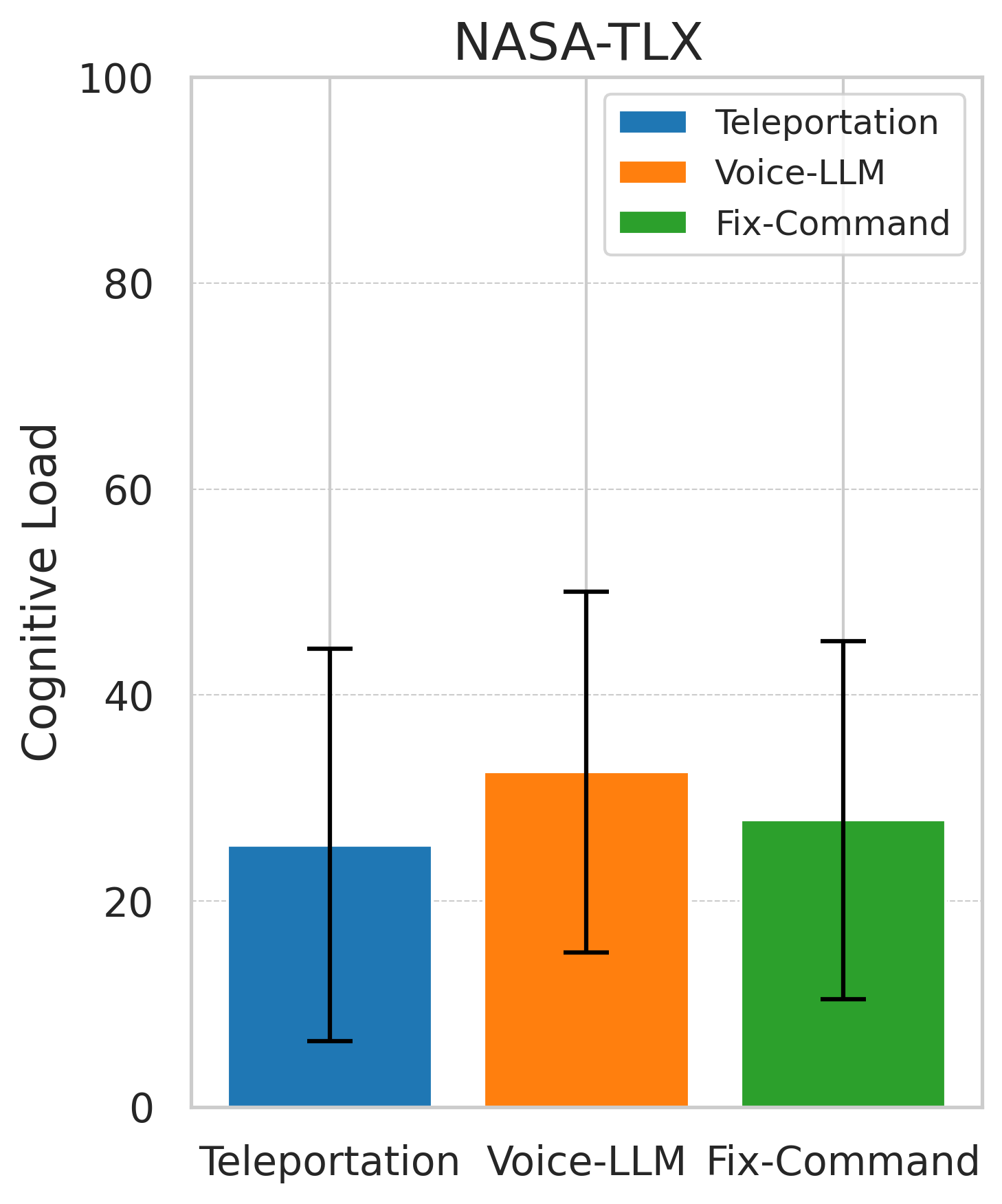}
        \caption{Results for NASA-TLX Scores.}
        \Description{Bar plots showing mean NASA Task Load Index (NASA-TLX) scores across the three navigation conditions. Error bars indicate variability across participants. Teleportation results in the lowest perceived cognitive load, while Voice-LLM and Fixed-Command show moderately higher workload levels.}
        \label{fig:nasatlx_scores}
    \end{minipage}%
    \hfill
    \begin{minipage}[t]{0.45\columnwidth}
        \centering
        \includegraphics[width=0.6\linewidth]{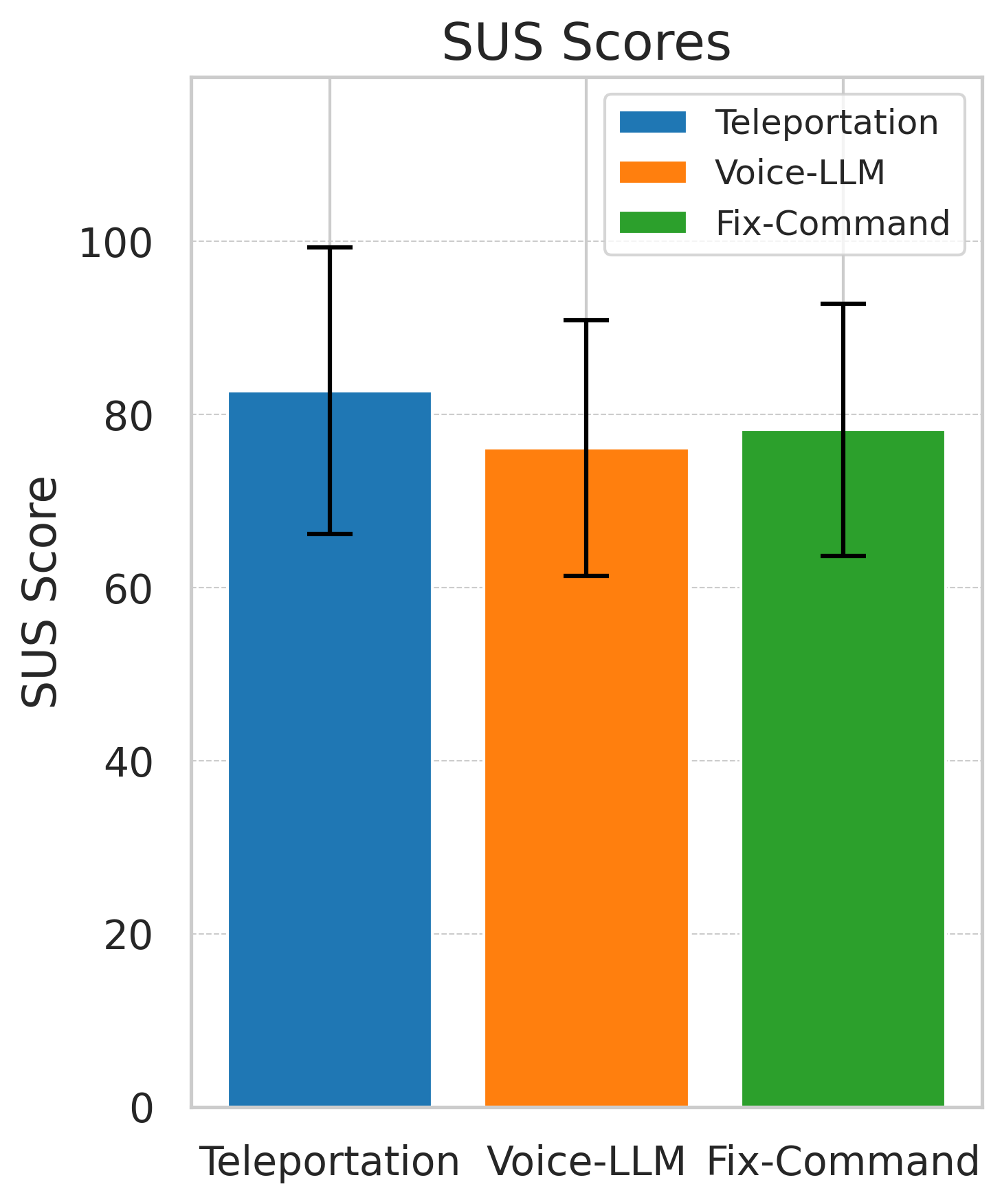}
        \caption{Results for SUS Scores.}
        \Description{Bar plots showing mean System Usability Scale (SUS) scores for the three navigation conditions. Error bars indicate variability across participants. All conditions receive high usability ratings, with Teleportation showing slightly higher scores than Voice-LLM and Fixed-Command.}
        \label{fig:sus_score}
    \end{minipage}
\end{figure}

The NASA-TLX results, illustrated in Figure~\ref{fig:nasatlx_scores} \revised{(error bars denote $\pm$1 SD across participants)}, present the perceived cognitive load associated with each locomotion technique. The score range for NASA-TLX is from 0 to 100. In general, participants did not report experiencing substantial cognitive load during the experiments. The teleportation condition yielded the lowest average score ($M = 25.44$, $SD = 19.07$), indicating the least cognitive demand among the three methods. In contrast, the LLM-driven locomotion method showed the highest average cognitive load ($M = 32.53$, $SD = 17.53$), while the voice-based steering method fell in between ($M = 27.86$, $SD = 17.39$). Despite these slight differences in mean values, statistical analysis revealed no significant differences between the distributions across conditions ($p = .918$).

\begin{table}[ht]
\small
\centering
\caption{Classification performance for predicting locomotion condition under leave-one-participant-out cross-validation (LOOCV). Final performance values are computed over the pooled out-of-fold predictions across all 60 folds.}
\label{tab:model_acc}
\revised{\Description{Table listing the classification accuracy of five models for predicting the three-class locomotion condition under leave-one-participant-out cross-validation. Logistic Regression L1 reaches 0.700 and Logistic Regression L2 reaches the best value 0.767, followed by Linear SVM 0.750, Ridge Regression 0.733, and LDA 0.683. The three-class chance level is 0.333.}}
\begin{tabular}{@{}p{6.2cm} p{2.5cm}@{}}
\toprule
\textbf{Model} & \textbf{Accuracy} \\
\midrule
\revised{Logistic Regression L1 ($C=3$)} & \revised{0.700} \\
\revised{Logistic Regression L2 ($C=3$)} & \revised{\textbf{0.767}} \\
Linear SVM, hinge ($C=2$) & 0.750 \\
\revised{Ridge Regression} ($\alpha=5$, SVD) & 0.733 \\
LDA (shrinkage = auto) & 0.683 \\
\bottomrule
\end{tabular}
\end{table}

\subsubsection{SUS Questionnaire}
The SUS results, shown in Figure~\ref{fig:sus_score} \revised{(error bars denote $\pm$1 SD across participants)}, reflect perceived usability. While there was no statistically significant difference between the techniques ($p = .383$), teleportation, the baseline method, received the highest usability score ($M = 82.75$, $SD = 16.54$), indicating a high level of user satisfaction and ease of use. Both the LLM-driven ($M = 76.13$, $SD = 14.77$) and voice-based steering conditions ($M = 78.25$, $SD = 14.58$) scored slightly lower. These results suggest generally good usability for both as hands-free alternatives, though less favorable than teleportation.

\subsection{Model} 
We conducted an exploratory analysis to evaluate the performance of various machine learning models and identify key eye-tracking features associated with the locomotion method. The dependent variable was the locomotion condition assigned to each participant, with features aggregated per participant by averaging across 3-second eye-tracking windows. We used leave-one-participant-out cross-validation across all $60$ participants. Since all three locomotion conditions had equal representation ($20$ samples each, $33.33\%$), the baseline accuracy obtained by predicting any single class was $0.333$, equal to the three-class chance level. We summarize the classification accuracies of each model in Table~\ref{tab:model_acc}. Among these, Logistic Regression L2 ($C = 3$) achieved the highest performance, with an accuracy of $0.767$, followed by Linear SVM, Ridge, LogReg L1, and LDA. The confusion
matrix and class-wise performance are reported in Appendix~\ref{app:model}.

\begin{figure}[ht]
    \centering
    \includegraphics[width=0.55\columnwidth]{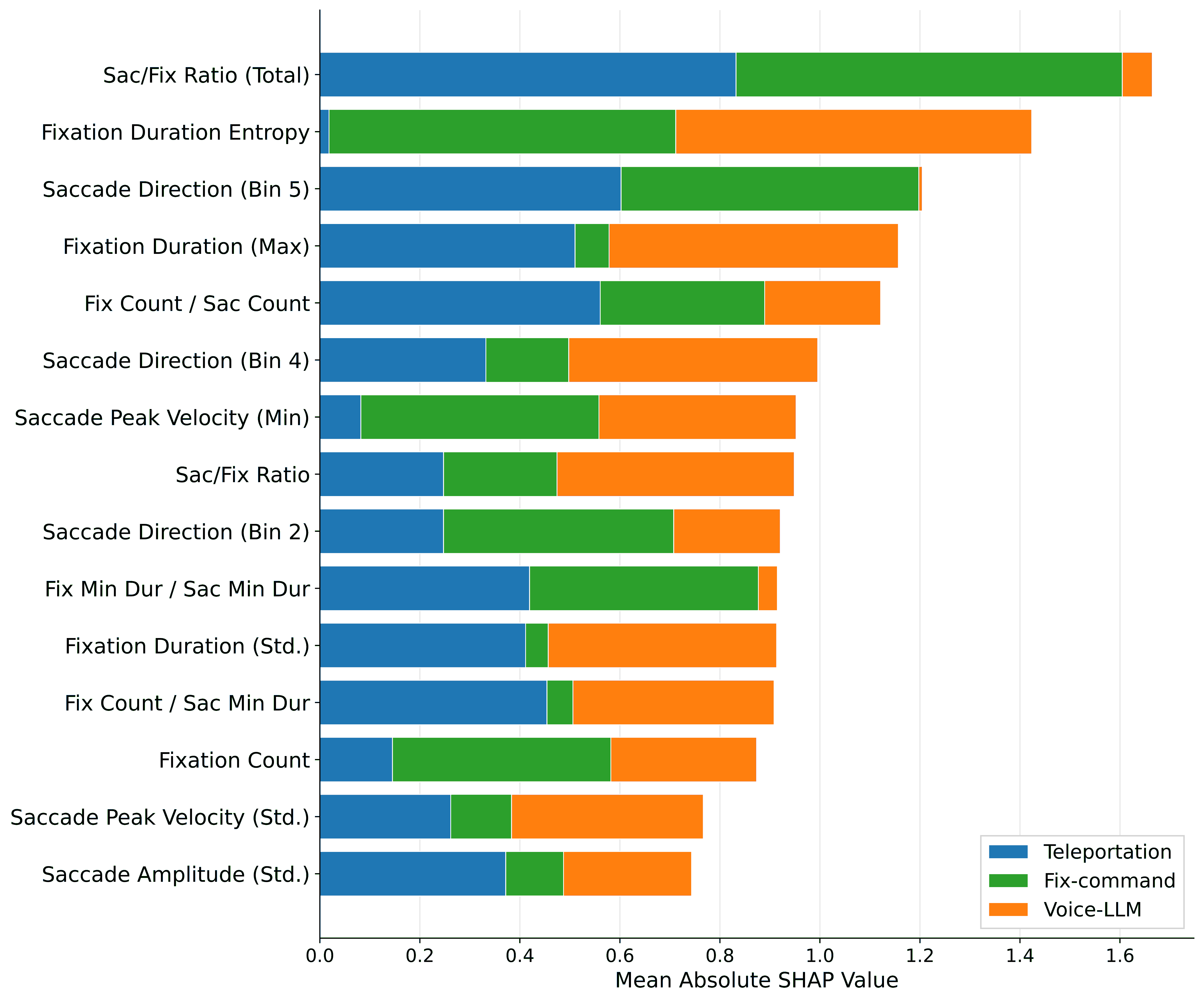}
\caption{SHAP summary plot showing the 15 most important eye-tracking features for predicting locomotion condition with the Logistic Regression L2 classifier.}
\Description{Stacked bar chart of mean absolute SHAP values illustrating the relative importance of eye-tracking features in the Logistic Regression L2 classifier. Feature contributions are decomposed by locomotion condition (Teleportation, Fixed-Command, and Voice-LLM), highlighting how different interaction techniques influence the model predictions. Higher values indicate greater impact on classification performance.}
    \label{fig:shap_summary}
\end{figure}

To interpret the effect of locomotion techniques on eye movement characteristics, we used SHAP~\cite{lundberg2017unified} to evaluate the contribution of individual features within the best-performing Logistic Regression L2 ($C=3$) classifier. The SHAP summary plot is shown in Figure~\ref{fig:shap_summary}, illustrating the per-class mean absolute SHAP value of each feature, aggregated across teleportation, voice-based steering through fixed commands, and LLM-driven locomotion conditions. According to the SHAP analysis, the saccade-to-fixation duration ratio, fixation duration entropy, and the saccade direction histogram~\cite{tatler2008systematic} emerged as the most important features for distinguishing between locomotion techniques. These features are commonly linked to cognitive load~\cite{2025.EDM.long-papers.9}, visual attention~\cite{shiferaw2019review}, engagement~\cite{holmqvist2011eye}, and visual search behavior~\cite{tatler2008systematic}, suggesting that user interaction patterns varied across conditions. Additionally, the maximum fixation duration and the fixation-to-saccade count ratio substantially contributed to the model predictions, indicating further differences in attentional strategies. Other important features included the minimum and variability of saccade peak velocity, fixation duration variability, fixation count, and saccade amplitude variability. Overall, the results indicate that saccade-to-fixation timing balance, fixation duration variability, and saccade directional patterns collectively shape the gaze responses associated with each locomotion technique, reflecting distinct patterns of cognitive processing and user engagement in VR.

\section{Discussion}
We investigated three locomotion techniques within a VR environment: teleportation, fixed voice command-based locomotion and context-aware LLM-driven locomotion. While teleportation is used as a baseline condition, the primary focus was to evaluate the effectiveness of the two hands-free techniques in terms of usability, user comfort, cognitive demand, and immersive experience.

\subsection{Performance and User Experience}
Teleportation, while the fastest method, relies on hand controllers, making it unsuitable for hands-free scenarios. Among the hands-free methods, voice-based steering initially enabled faster navigation to the first target, while the LLM-driven method showed descriptively lower completion time in the second task segment, as shown in Section \ref{sec:performance_analysis}. Although this difference was not statistically significant, it may suggest that users can adapt to the natural language system and use its flexibility after brief exposure, despite being unfamiliar with this form of locomotion. This is consistent with prior voice-based locomotion work, which suggests that simple fixed commands can support efficient real-time directional control, whereas semantic or natural-language-oriented interfaces can provide greater flexibility when users refer to landmarks, scene objects, or task context~\cite{hombeck2023tell,ferracani2017natural,calandra2022comparison}.

Across usability, cognitive load, cybersickness, and presence we did not observe statistically significant differences between conditions. Using established SUS adjective ratings~\cite{bangor2009determining}, teleportation received the highest usability score indicating “excellent” usability. Both voice-based methods received scores between 75 and 80, reflecting “good” usability. These results demonstrate that both hands-free techniques were generally well-received by users, even though they did not match the efficiency or ease of teleportation.

The LLM-driven method, which enables instant transport to target locations, showed descriptively lower nausea and vestibular discomfort, but we found no evidence of differences in CSQ-VR scores across conditions. This aligns with prior work suggesting that locomotion techniques involving continuous visual motion can increase cybersickness risk, whereas instant teleportation can reduce sensory conflict by avoiding the visual transition between locations~\cite{bozgeyikli2016point,mayor2019comparative,biswas2024you}. Although all participants reached the same targets, those using teleportation completed the task much faster, resulting in substantially less exposure to the virtual environment---\revised{nearly three times shorter} than the other conditions. This reduced exposure duration likely contributed to the lower overall cybersickness scores in the teleportation condition, consistent with prior findings that cybersickness severity can increase with longer VR exposure~\cite{biswas2024you}. Interestingly, the pattern reversed descriptively for oculomotor discomfort, with voice-based steering scoring slightly lower. This may indicate that abrupt viewpoint changes in teleportation affect oculomotor symptoms differently from nausea or vestibular discomfort, although no significant differences were observed.

Teleportation required the least cognitive effort, while the LLM-driven method showed slightly higher demand than voice-based steering, although these differences were not statistically significant. A similar trend was observed in the pupil diameter measurements. Since pupil diameter is commonly associated with mental effort, arousal, and task engagement~\cite{beatty1982task,hopstaken2015window}, the increase in the LLM-driven condition may reflect the mental effort required to plan contextually relevant and efficient commands, which is not expected in teleportation or in command-based methods that rely on a simple and repeatable set of inputs. However, it may also promote deeper cognitive engagement and encourage more active exploration of the immersive virtual environment, potentially contributing to the heightened sense of presence reported in the LLM-driven locomotion condition.  Additionally, although free-form commands may initially impose a higher cognitive load, this demand is likely to decrease with experience, which can be further investigated through longitudinal studies. Despite the slightly higher cognitive load associated with contextual commands, participants using LLM-driven locomotion reported the highest levels of general and spatial presence. Actively engaging with the environment to provide context-aware instructions likely strengthened their sense of being “present” compared to voice-based steering. In contrast, voice-based steering received the highest realism scores, possibly because its continuous walking pattern more closely mimics real-world locomotion.

In addition to the standardized questionnaires, participants provided positive qualitative feedback\revised{ (open-ended verbal comments collected during and immediately after the session)} across all locomotion strategies, although each evaluated only one method. Teleportation was commonly described as intuitive and easy to adapt to, contributing to a smooth experience. Voice-based steering was perceived as natural due to its similarity to real-world walking and the feeling of continuous movement throughout navigation. However, participants occasionally attempted to use command variations beyond the predefined set, which sometimes led to unrecognized inputs. This mirrors earlier observations that grammar-based voice interfaces can be straightforward to learn and use, but may restrict phrasing flexibility when interaction is limited to predefined commands or utterance patterns~\cite{ferracani2017natural,calandra2022comparison,hombeck2023tell}. The LLM-driven approach received particularly positive feedback. Many participants described it as highly intuitive, engaging, and enjoyable to use. In fact, during the demo scene, some users experimented with different phrasings, suggesting that the open-ended input style encouraged curiosity, exploration, and deeper engagement with the virtual environment. We interpret this as an important qualitative benefit of context-aware language input: instead of only issuing directional controls, users could test how the system understood the scene and formulate navigation goals in their own words. This suggests that LLM-driven locomotion may support more exploratory forms of interaction, which should be studied directly in future work. However, there were a few instances where the system failed to understand commands, mainly due to limitations in the speech-to-text module.

\subsection{Cognitive and Visual Attention}
Voice-based steering resulted in significantly higher fixation rates, suggesting more frequent visual sampling during continuous navigation. This pattern may be attributed to the nature of constant movement, where users need to monitor their direction and search the environment while moving, often resulting in shorter and more frequent fixations. This interpretation is consistent with steering-based travel work, where users continuously regulate direction and move through the scene rather than selecting a discrete destination~\cite{christou2017steering,hombeck2023tell}. Similarly, mean fixation duration can reflect how stably attention is allocated to visual information. Teleportation resulted in the longest fixations, followed by the LLM-driven method, while the voice-based steering condition showed the shortest durations. These eye-movement patterns, particularly less frequent but longer fixations in the teleportation and LLM-driven conditions, suggest that users may have planned their actions more deliberately and were generally aware of their intended destinations. 

Mean saccade durations were significantly shorter for LLM-driven locomotion, suggesting more efficient and goal-directed gaze behavior, as saccade measures are commonly used to characterize visual search and attention allocation~\cite{holmqvist2011eye,eckstein2017beyond}. Additionally, while pupil diameter is primarily considered an indicator of cognitive load, increased pupil size may also reflect heightened engagement or arousal within the virtual setting~\cite{beatty1982task,hopstaken2015window}. This finding is further supported by the higher spatial and general presence scores in the LLM-driven condition, suggesting that natural language interaction facilitated deeper cognitive engagement and immersion.

Classification models demonstrate the feasibility of using eye-tracking features to infer user interaction mode in real time, as these features varied across locomotion techniques. SHAP analysis indicated that the saccade-to-fixation duration ratio, fixation duration entropy and variability, saccade direction histograms, maximum fixation duration, and the fixation-to-saccade count ratio were key indicators for distinguishing user behavior.

Together, these results suggest that locomotion techniques induce distinguishable gaze-movement signatures, primarily through differences in saccadic speed, temporal variability, and gaze-transition behavior. Rather than relying on a single eye-tracking marker, the classifier appears to capture a combination of dynamic oculomotor features that reflect how users visually explore and control movement under each interaction technique. These findings also support the use of eye-tracking features as objective signals for adaptive VR systems, consistent with prior work on cognitive-load estimation and adaptive interaction in VR~\cite{zagermann2016measuring,bozkir2019person,bozkir2019assessment}.

\subsection{Accessibility and Practical Considerations}
The LLM-driven locomotion method provides a highly accessible alternative for users with disabilities or temporary impairments, as it does not require additional hardware such as handheld controllers, and matches teleportation and fixed-command methods in usability and comfort. With its inherent multilingual support, users can interact naturally in their own languages without relying on language-specific fixed commands. Recent LLM-based XR work similarly points to language models as a way to support flexible and inclusive interaction when user intent can be grounded in context~\cite{bozkir2024embedding,skyba2024towards}. In many practical virtual environments, existing scene information is already sufficiently descriptive to support this approach with minimal additional setup, making the method adaptable and inclusive. However, not all environments may provide sufficiently structured scene information for reliable interpretation; therefore, additional annotations may still be required. Such flexibility is especially valuable in settings like medical training, education, and remote collaboration, where accessibility and ease of use are critical as VR adoption grows.

At the same time, several practical challenges should be considered. Handling variability in user prompts requires careful prompt design and contextual grounding, particularly for real-time proper responses. In our implementation, the system prompt incorporated essential scene context (see Figure~\ref{fig:teaser}), enabling the LLM to interpret intent accurately and generate only the target coordinates. Outputs are short, and responses average 1.5 seconds without disrupting interaction or causing discomfort, as the movement is not continuous. Latency can be further reduced by using smaller LLMs and lighter speech-to-text models, although prior XR work also notes that robust spatial language understanding remains challenging~\cite{skyba2024towards}.

Another challenge is the inherently dynamic and sometimes unpredictable nature of LLM outputs. Ensuring consistent behavior in real-world scenarios requires error-handling mechanisms, such as constraining movement ranges, validating coordinates, or ignoring ambiguous outputs. 
In our implementation, we used ChatGPT-4o~\cite{gpt4o} because it provided reliable context interpretation and consistently formatted coordinate outputs during development. Although smaller LLMs could lower latency, they may require more careful prompt design and validation to maintain comparable reliability.
Additionally, our approach represents the scene as text, enabling real-time processing but potentially overlooking spatial layouts. As scene complexity increases, larger prompts may introduce latency and reduce robustness. Nevertheless, LLMs can interpret structured XR inputs and natural-language commands when prompts provide sufficient scene and task context, although robustness may decrease in more complex spatial scenarios~\cite{skyba2024towards,buldu2025cuify}.

\subsection{Limitations}
To support replication, we provide all experimental details; however, exact reproducibility cannot be guaranteed due to the non-deterministic nature of language models, although using the same model and configuration is expected to yield comparable results. We used a trigger button in both voice-based methods, which required a controller to enable an objective comparison and minimize the effect of design preferences. Additionally, we used a between-subjects design to avoid learning and familiarity effects with the environment; however, participants rated only their assigned technique, which limits direct comparisons across locomotion methods.

The evaluated conditions compare practical locomotion techniques rather than fully isolating input modality, movement type, and language understanding. Consequently, differences between the LLM-driven and fixed-command conditions cannot be attributed only to the LLM, because they also differ in teleportation versus continuous steering and in ASR pipeline. A controller-steering condition could further isolate movement type, but it would not address the hands-free motivation of the work and would require a larger factorial design. Future work should include stronger hands-free baselines, such as head/gaze-based pointing with voice or other confirmation modalities, to isolate the added value of free-form LLM interpretation.


The current task used a simplified town layout with visible targets. Moreover, the approach still depends on available scene metadata, as objects need descriptive names or tags for semantic interpretation. However, in practical VR or simulation environments, such metadata is often already available through object names, tags, or scene hierarchies, meaning that our method does not necessarily require additional manual annotation. Thus, our results support feasibility in a controlled setting, but they do not yet demonstrate performance in complex, dynamic, or entirely unannotated environments. More complex and dynamic environments should be evaluated in future work.

\revised{Finally, because the LLM-driven technique transmits spoken commands and runtime scene context to a language model and relies on eye tracking for analysis, it raises privacy considerations around gaze, speech, and interaction data~\cite{bozkir2023eye,abdrabou2025gaze}. Privacy-preserving processing, such as on-device or local language and speech models, is therefore an important direction for future work.}

\section{Conclusion}
In this work, we proposed an LLM-driven, hands-free locomotion technique for VR that uses natural language instructions and contextual scene information. To assess its feasibility, we compared three locomotion techniques: controller-based teleportation, voice-based steering through fixed commands, and our LLM-driven approach. We combined eye-tracking analysis with standardized questionnaires, including the SUS, IPQ, CSQ-VR, and NASA-TLX, to assess visual behavior, usability, presence, cybersickness, and cognitive workload. Compared with fixed-command voice steering, the LLM-driven method offered a more flexible selection-based alternative: users could issue free-form, context-aware commands instead of relying on a small command vocabulary, and the instant movement design avoided the continuous visual motion of steering. This was reflected in descriptively lower nausea and vestibular discomfort, lower fixation rates than voice steering, and significantly shorter mean saccade durations. Compared with controller-based teleportation, the LLM-driven method did not match the speed of manual teleportation, but it achieved comparable subjective usability, comfort, and presence while reducing reliance on handheld controller pointing for destination selection. Overall, our results demonstrate that LLM-driven locomotion is a promising alternative for intuitive, accessible, and immersive hands-free navigation in VR environments. Future research should investigate personalization techniques, such as adapting LLM behavior based on user preferences or prior interactions, to enhance predictability and usability.
\bibliographystyle{ACM-Reference-Format}
 \bibliography{references}

@article{biswas2024you,
  title={“Are you feeling sick?”--A systematic literature review of cybersickness in virtual reality},
  author={Biswas, Nilotpal and Mukherjee, Anamitra and Bhattacharya, Samit},
  journal={ACM Computing Surveys},
  volume={56},
  number={11},
  pages={1--38},
  year={2024},
  publisher={ACM New York, NY}
}

@article{eckstein2017beyond,
  title={Beyond eye gaze: What else can eyetracking reveal about cognition and cognitive development?},
  author={Eckstein, Maria K and Guerra-Carrillo, Bel{\'e}n and Singley, Alison T Miller and Bunge, Silvia A},
  journal={Developmental cognitive neuroscience},
  volume={25},
  pages={69--91},
  year={2017},
  publisher={Elsevier}
}

@inproceedings{gao2022eye,
  title={Eye-Tracking-Based Prediction of User Experience in VR Locomotion Using Machine Learning},
  author={Gao, Hong and Kasneci, Enkelejda},
  booktitle={Computer Graphics Forum},
  volume={41},
  number={7},
  pages={589--599},
  year={2022},
  organization={Wiley Online Library}
}

@incollection{hart1988development,
  title={Development of NASA-TLX (Task Load Index): Results of empirical and theoretical research},
  author={Hart, Sandra G and Staveland, Lowell E},
  booktitle={Advances in psychology},
  volume={52},
  year={1988},
  publisher={Elsevier}
}

@inproceedings{kourtesis2023cybersickness,
  title={Cybersickness in virtual reality questionnaire (csq-vr): A validation and comparison against ssq and vrsq},
  author={Kourtesis, Panagiotis and Linnell, Josie and Amir, Rayaan and Argelaguet, Ferran and MacPherson, Sarah E},
  booktitle={Virtual Worlds},
  year={2023},
}

@article{schubert2001experience,
  title={The experience of presence: Factor analytic insights},
  author={Schubert, Thomas and Friedmann, Frank and Regenbrecht, Holger},
  journal={Presence: Teleoperators \& Virtual Environments},
  volume={10},
  number={3},
  year={2001},
  publisher={MIT Press}
}

@article{lundberg2017unified,
  title={A unified approach to interpreting model predictions},
  author={Lundberg, Scott M and Lee, Su-In},
  journal={Advances in neural information processing systems},
  volume={30},
  year={2017}
}

@article{clifton2020effects,
  title={Effects of steering locomotion and teleporting on cybersickness and presence in HMD-based virtual reality},
  author={Clifton, Jeremy and Palmisano, Stephen},
  journal={Virtual Reality},
  volume={24},
  number={3},
  year={2020},
  publisher={Springer}
}

@inproceedings{buldu2025cuify,
  title={{CUIfy the XR}: An Open-Source Package to Embed LLM-powered Conversational Agents in {XR}},
  author={Buldu, Kadir Burak and {\"O}zdel, S{\"u}leyman and Lau, Ka Hei Carrie and Wang, Mengdi and Saad, Daniel and Sch{\"o}nborn, Sofie and Boch, Auxane and Kasneci, Enkelejda and Bozkir, Efe},
  booktitle={2025 IEEE International Conference on Artificial Intelligence and eXtended and Virtual Reality (AIxVR)},
  year={2025},
  organization={IEEE}
}

@article{bates1992virtual,
  title={Virtual reality, art, and entertainment},
  author={Bates, Joseph},
  journal={Presence: Teleoperators \& Virtual Environments},
  volume={1},
  number={1},
  year={1992},
  publisher={MIT Press}
}

@article{hopstaken2015window,
  title={The window of my eyes: Task disengagement and mental fatigue covary with pupil dynamics},
  author={Hopstaken, Jesper F and Van Der Linden, Dimitri and Bakker, Arnold B and Kompier, Michiel AJ},
  journal={Biological psychology},
  volume={110},
  year={2015},
  publisher={Elsevier}
}

@article{benedetto2011driver,
  title={Driver workload and eye blink duration},
  author={Benedetto, Simone and Pedrotti, Marco and Minin, Luca and Baccino, Thierry and Re, Alessandra and Montanari, Roberto},
  journal={Transportation research part F: traffic psychology and behaviour},
  volume={14},
  number={3},
  year={2011},
  publisher={Elsevier}
}

@article{just1980theory,
  title={A theory of reading: from eye fixations to comprehension.},
  author={Just, Marcel A and Carpenter, Patricia A},
  journal={Psychological review},
  volume={87},
  number={4},
  year={1980},
  publisher={American Psychological Association}
}

@inproceedings{liu2021force,
  title={Force-based foot gesture navigation in virtual reality},
  author={Liu, Siyi and Lee, Gun and Li, Yi and Piumsomboon, Thammathip and Ens, Barrett},
  booktitle={Proceedings of the 27th ACM Symposium on Virtual Reality Software and Technology},
  year={2021}
}

@article{prithul2021teleportation,
  title={Teleportation in virtual reality; a mini-review},
  author={Prithul, Aniruddha and Adhanom, Isayas Berhe and Folmer, Eelke},
  journal={Frontiers in Virtual Reality},
  volume={2},
  year={2021},
  publisher={Frontiers Media SA}
}

@inproceedings{zagermann2016measuring,
  title={Measuring cognitive load using eye tracking technology in visual computing},
  author={Zagermann, Johannes and Pfeil, Ulrike and Reiterer, Harald},
  booktitle={Proceedings of the sixth workshop on beyond time and errors on novel evaluation methods for visualization},
  year={2016}
}

@book{holmqvist2011eye,
  title={Eye tracking: A comprehensive guide to methods and measures},
  author={Holmqvist, Kenneth and Nystr{\"o}m, Marcus and Andersson, Richard and Dewhurst, Richard and Jarodzka, Halszka and Van de Weijer, Joost},
  year={2011},
  publisher={OUP Oxford}
}

@article{bruder2012redirecting,
  title={Redirecting walking and driving for natural navigation in immersive virtual environments},
  author={Bruder, Gerd and Interrante, Victoria and Phillips, Lane and Steinicke, Frank},
  journal={IEEE transactions on visualization and computer graphics},
  volume={18},
  number={4},
  year={2012},
  publisher={IEEE}
}

@inproceedings{bowman1997travel,
  title={Travel in immersive virtual environments: An evaluation of viewpoint motion control techniques},
  author={Bowman, Doug A and Koller, David and Hodges, Larry F},
  booktitle={Proceedings of IEEE 1997 Annual International Symposium on Virtual Reality},
  year={1997},
  organization={IEEE}
}

@inproceedings{qian2018look,
  title={Look to go: An empirical evaluation of eye-based travel in virtual reality},
  author={Qian, Yuan Yuan and Teather, Robert J},
  booktitle={Proceedings of the 2018 ACM Symposium on Spatial User Interaction},
  year={2018}
}

@inproceedings{sargunam2018evaluating,
  title={Evaluating joystick control for view rotation in virtual reality with continuous turning, discrete turning, and field-of-view reduction},
  author={Sargunam, Shyam Prathish and Ragan, Eric D},
  booktitle={Proceedings of the 3rd International Workshop on Interactive and Spatial Computing},
  year={2018}
}

@inproceedings{wilson2016vr,
  title={VR locomotion: walking> walking in place> arm swinging},
  author={Wilson, Preston Tunnell and Kalescky, William and MacLaughlin, Ansel and Williams, Betsy},
  booktitle={Proceedings of the 15th ACM SIGGRAPH conference on virtual-reality continuum and its applications in industry-volume 1},
  year={2016}
}

@inproceedings{ferracani2017natural,
  title={Natural experiences in museums through virtual reality and voice commands},
  author={Ferracani, Andrea and Faustino, Marco and Giannini, Gabriele Xavier and Landucci, Lea and Del Bimbo, Alberto},
  booktitle={Proceedings of the 25th ACM international conference on Multimedia},
  year={2017}
}

@misc{shmyrev2020vosk,
  title={Vosk speech recognition toolkit: offline speech recognition API for android, iOS, Raspberry Pi and servers with Python, Java, C\# and Node},
  author={Shmyrev, NV and Team, Vosk Core},
  year={2020},
  publisher={GitHub repository}
}

@inproceedings{mardanbegi2019eyemrtk,
  title={Eyemrtk: A toolkit for developing eye gaze interactive applications in virtual and augmented reality},
  author={Mardanbegi, Diako and Pfeiffer, Thies},
  booktitle={Proceedings of the 11th ACM Symposium on Eye Tracking Research \& Applications},
  year={2019}
}

@article{jacob1991use,
  title={The use of eye movements in human-computer interaction techniques: what you look at is what you get},
  author={Jacob, Robert JK},
  journal={ACM Transactions on Information Systems (TOIS)},
  volume={9},
  number={2},
  year={1991},
  publisher={ACM}
}

@manual{OculusBestPractices,
  title        = {Oculus Best Practices},
  author       = {{Oculus VR, LLC}},
  year         = {2017},
  url          = {https://www.cs.sjtu.edu.cn/~shengbin/course/vr/misc/oculus_bestpractices.pdf},
}

@inproceedings{blattgerste2018advantages,
  title={Advantages of eye-gaze over head-gaze-based selection in virtual and augmented reality under varying field of views},
  author={Blattgerste, Jonas and Renner, Patrick and Pfeiffer, Thies},
  booktitle={Proceedings of the workshop on communication by gaze interaction},
  year={2018}
}

@inproceedings{sra2018breathvr,
  title={Breathvr: Leveraging breathing as a directly controlled interface for virtual reality games},
  author={Sra, Misha and Xu, Xuhai and Maes, Pattie},
booktitle = {Proceedings of the 2018 {CHI} Conference on Human Factors in Computing Systems},
year={2018}
}

@article{alfaro2018scientific,
  title={Scientific articles exploration system model based in immersive virtual reality and natural language processing techniques},
  author={Alfaro, Luis and Linares, Ricardo and Herrera, Jose},
  journal={International Journal of Advanced Computer Science and Applications},
  volume={9},
  number={7},
  year={2018},
  publisher={Science and Information (SAI) Organization Limited}
}

@article{caggianese2020freehand,
  title={Freehand-steering locomotion techniques for immersive virtual environments: A comparative evaluation},
  author={Caggianese, Giuseppe and Capece, Nicola and Erra, Ugo and Gallo, Luigi and Rinaldi, Michele},
  journal={International Journal of Human--Computer Interaction},
  volume={36},
  number={18},
  pages={1734--1755},
  year={2020},
  publisher={Taylor \& Francis}
}

@inproceedings{christou2017steering,
  title={Steering versus teleport locomotion for head mounted displays},
  author={Christou, Chris G and Aristidou, Poppy},
  booktitle={Augmented Reality, Virtual Reality, and Computer Graphics: 4th International Conference, AVR 2017, Ugento, Italy, June 12-15, 2017, Proceedings, Part II 4},
  year={2017},
  organization={Springer}
}

@article{hepperle20192d,
  title={2D, 3D or speech? A case study on which user interface is preferable for what kind of object interaction in immersive virtual reality},
  author={Hepperle, Daniel and Wei{\ss}, Yannick and Siess, Andreas and W{\"o}lfel, Matthias},
  journal={Computers \& Graphics},
  volume={82},
  year={2019},
  publisher={Elsevier}
}

@inproceedings{schroeder2017presence,
  title={Presence and usability do not directly predict procedural recall in virtual reality training},
  author={Schroeder, Bradford L and Bailey, Shannon KT and Johnson, Cheryl I and Gonzalez-Holland, Emily},
  booktitle={International conference on human-computer interaction},
  year={2017},
  organization={Springer}
}

@inproceedings{gallo2008toward,
  title={Toward a natural interface to virtual medical imaging environments},
  author={Gallo, Luigi and De Pietro, Giuseppe and Coronato, Antonio and Marra, Ivana},
  booktitle={Proceedings of the working conference on Advanced visual interfaces},
  year={2008}
}

@inproceedings{9995155,
  author={Meng, Xuanru and Xu, Wenge and Liang, Hai-Ning},
  booktitle={2022 IEEE International Symposium on Mixed and Augmented Reality (ISMAR)}, 
  title={An Exploration of Hands-free Text Selection for Virtual Reality Head-Mounted Displays}, 
  year={2022},
  doi={10.1109/ISMAR55827.2022.00021}}

@article{rahimi2018scene,
  title={Scene transitions and teleportation in virtual reality and the implications for spatial awareness and sickness},
  author={Rahimi, Kasra and Banigan, Colin and Ragan, Eric D},
  journal={IEEE transactions on visualization and computer graphics},
  volume={26},
  number={6},
  year={2020},
  publisher={IEEE}
}

@inproceedings{paulo2020improving,
  title={Improving camera travel for immersive colonography},
  author={Paulo, Soraia F and Medeiros, Daniel and Borges, Pedro B and Jorge, Joaquim and Lopes, Daniel S},
  booktitle={2020 IEEE Conference on Virtual Reality and 3D User Interfaces Abstracts and Workshops (VRW)},
  year={2020},
  organization={IEEE}
}

@article{raees2019ven,
  title={VEN-3DVE: vision based egocentric navigation for 3D virtual environments},
  author={Raees, Muhammad and Ullah, Sehat and Rahman, Sami Ur},
  journal={International Journal on Interactive Design and Manufacturing (IJIDeM)},
  volume={13},
  number={1},
  year={2019},
  publisher={Springer}
}

@article{cho2017multi,
  title={Multi-scale 7DOF view adjustment},
  author={Cho, Isaac and Li, Jialei and Wartell, Zachary},
  journal={IEEE transactions on visualization and computer graphics},
  volume={24},
  number={3},
  year={2017},
  publisher={IEEE}
}

@inproceedings{medeiros2016effects,
  title={Effects of speed and transitions on target-based travel techniques},
  author={Medeiros, Daniel and Cordeiro, Eduardo and Mendes, Daniel and Sousa, Maur{\'\i}cio and Raposo, Alberto and Ferreira, Alfredo and Jorge, Joaquim},
  booktitle={Proceedings of the 22Nd ACM Conference on Virtual Reality Software and Technology},
  year={2016}
}

@inproceedings{liu2018increasing,
  title={Increasing walking in VR using redirected teleportation},
  author={Liu, James and Parekh, Hirav and Al-Zayer, Majed and Folmer, Eelke},
  booktitle={Proceedings of the 31st annual ACM symposium on user interface software and technology},
  year={2018}
}

@inproceedings{bozgeyikli2016point,
  title={Point \& teleport locomotion technique for virtual reality},
  author={Bozgeyikli, Evren and Raij, Andrew and Katkoori, Srinivas and Dubey, Rajiv},
  booktitle={Proceedings of the 2016 annual symposium on computer-human interaction in play},
  year={2016}
}

@inproceedings{feasel2008llcm,
  title={LLCM-WIP: Low-latency, continuous-motion walking-in-place},
  author={Feasel, Jeff and Whitton, Mary C and Wendt, Jeremy D},
  booktitle={2008 IEEE symposium on 3D user interfaces},
  year={2008},
  organization={IEEE}
}

@inproceedings{calandra2019usability,
  title={On the usability of consumer locomotion techniques in serious games: Comparing arm swinging, treadmills and walk-in-place},
  author={Calandra, Davide and Lamberti, Fabrizio and Migliorini, Massimo},
  booktitle={2019 IEEE 9th International Conference on Consumer Electronics (ICCE-Berlin)},
  year={2019},
  organization={IEEE}
}

@inproceedings{tan2022understanding,
  title={Understanding user experiences across VR Walking-in-Place locomotion methods},
  author={Tan, Chek Tien and Foo, Leon Cewei and Yeo, Adriel and Lee, Jeannie Su Ann and Wan, Edmund and Kok, Xiao-Feng Kenan and Rajendran, Megani},
  booktitle={Proceedings of the 2022 CHI Conference on Human Factors in Computing Systems},
  year={2022}
}

@article{wehden2021slippery,
  title={The slippery path to total presence: How omnidirectional virtual reality treadmills influence the gaming experience},
  author={Wehden, Lars-Ole and Reer, Felix and Janzik, Robin and Tang, Wai Yen and Quandt, Thorsten},
  journal={Media and Communication},
  volume={9},
  number={1},
  year={2021}
}

@inproceedings{langbehn2018evaluation,
  title={Evaluation of locomotion techniques for room-scale vr: Joystick, teleportation, and redirected walking},
  author={Langbehn, Eike and Lubos, Paul and Steinicke, Frank},
  booktitle={Proceedings of the Virtual Reality International Conference-Laval Virtual},
  year={2018}
}

@inproceedings{martinez2022research,
  title={Research trends in virtual reality locomotion techniques},
  author={Martinez, Esteban Segarra and Wu, Annie S and McMahan, Ryan P},
  booktitle={2022 IEEE Conference on Virtual Reality and 3D User Interfaces (VR)},
  year={2022},
  organization={IEEE}
}

@article{anderton2025teleportation,
  title={From teleportation to climbing: A review of locomotion techniques in the most used commercial virtual reality applications},
  author={Anderton, Craig and Creed, Chris and Sarcar, Sayan and Theil, Arthur},
  journal={International Journal of Human--Computer Interaction},
  volume={41},
  number={4},
  year={2025},
  publisher={Taylor \& Francis}
}

@inproceedings{coomer2018evaluating,
  title={Evaluating the effects of four VR locomotion methods: joystick, arm-cycling, point-tugging, and teleporting},
  author={Coomer, Noah and Bullard, Sadler and Clinton, William and Williams-Sanders, Betsy},
  booktitle={Proceedings of the 15th ACM symposium on applied perception},
  year={2018}
}

@article{bozgeyikli2019locomotion,
  title={Locomotion in virtual reality for room scale tracked areas},
  author={Bozgeyikli, Evren and Raij, Andrew and Katkoori, Srinivas and Dubey, Rajiv},
  journal={International Journal of Human-Computer Studies},
  volume={122},
  year={2019},
  publisher={Elsevier}
}

@article{xie2021review,
  title={A review on virtual reality skill training applications},
  author={Xie, Biao and Liu, Huimin and Alghofaili, Rawan and Zhang, Yongqi and Jiang, Yeling and Lobo, Flavio Destri and Li, Changyang and Li, Wanwan and Huang, Haikun and Akdere, Mesut and others},
  journal={Frontiers in Virtual Reality},
  volume={2},
  year={2021},
  publisher={Frontiers Media SA}
}

@article{halbig2022opportunities,
  title={Opportunities and challenges of virtual reality in healthcare--a domain experts inquiry},
  author={Halbig, Andreas and Babu, Sooraj K and Gatter, Shirin and Latoschik, Marc Erich and Brukamp, Kirsten and Von Mammen, Sebastian},
  journal={Frontiers in Virtual Reality},
  volume={3},
  year={2022},
}

@article{mayor2019comparative,
  title={A comparative study of virtual reality methods of interaction and locomotion based on presence, cybersickness, and usability},
  author={Mayor, Jesus and Raya, Laura and Sanchez, Alberto},
  journal={IEEE Transactions on Emerging Topics in Computing},
  volume={9},
  number={3},
  year={2019},
  publisher={IEEE}
}

@inproceedings{calandra2022comparison,
  title={Comparison of hands-free speech-based navigation techniques for virtual reality training},
  author={Calandra, Davide and Prattic{\`o}, Filippo Gabriele and Lamberti, Fabrizio},
  booktitle={2022 IEEE 21st Mediterranean Electrotechnical Conference (MELECON)},
  year={2022},
  organization={IEEE}
}

@article{monteiro2021hands,
  title={Hands-free interaction in immersive virtual reality: A systematic review},
  author={Monteiro, Pedro and Gon{\c{c}}alves, Guilherme and Coelho, Hugo and Melo, Miguel and Bessa, Maximino},
  journal={IEEE Transactions on Visualization and Computer Graphics},
  volume={27},
  number={5},
  year={2021},
  publisher={IEEE}
}

@misc{igroup_ipq,
  author       = {Thomas Schubert and Frank Friedmann and Holger Regenbrecht},
  title        = {Igroup Presence Questionnaire (IPQ)},
  year         = {2001},
  howpublished = {\url{https://www.igroup.org/pq/ipq/index.php}},
  note         = {Accessed: 2025-03-21},
  institution  = {igroup.org – project consortium},
}

@article{brooke1996sus,
  title={SUS-A quick and dirty usability scale},
  author={Brooke, John and others},
  journal={Usability evaluation in industry},
  volume={189},
  number={194},
  year={1996},
  publisher={London, England.}
}

@inproceedings{hombeck2023tell,
  title={Tell me where to go: Voice-controlled hands-free locomotion for virtual reality systems},
  author={Hombeck, Jan and Voigt, Henrik and Heggemann, Timo and Datta, Rabi R and Lawonn, Kai},
  booktitle={2023 IEEE Conference Virtual Reality and 3D User Interfaces (VR)},
  year={2023},
  organization={IEEE}
}

@inproceedings{gao2021digital,
  title={Digital transformations of classrooms in virtual reality},
  author={Gao, Hong and Bozkir, Efe and Hasenbein, Lisa and Hahn, Jens-Uwe and G{\"o}llner, Richard and Kasneci, Enkelejda},
  booktitle={Proceedings of the 2021 CHI Conference on Human Factors in Computing Systems},
  year={2021}
}

@inproceedings{ansari2022implementing,
  title={Implementing virtual reality in entertainment industry},
  author={Ansari, Saniya Zubair Ahmed and Shukla, Vinod Kumar and Saxena, Komal and Filomeno, Bethoven},
  booktitle={Cyber Intelligence and Information Retrieval: Proceedings of CIIR 2021},
  year={2022},
  organization={Springer}
}

@misc{gpt4o,
  author    = {OpenAI},
  title     = {Hello GPT-4o},
  year      = {2024},
  url       = {https://openai.com/hello-gpt-4o/},
  note      = {Accessed: 2024-08-30}
}

@inproceedings{salvucci2000identifying,
  title={Identifying fixations and saccades in eye-tracking protocols},
  author={Salvucci, Dario D and Goldberg, Joseph H},
  booktitle={Proceedings of the 2000 symposium on Eye tracking research \& applications},
  year={2000}
}

@misc{varjo_xr3,
  author       = {{Varjo Technologies}},
  title        = {{Varjo XR-3}: Mixed Reality Headset},
  year         = 2024,
  url          = {https://varjo.com/xr-headsets/},
  note         = {Accessed: 2024-09-12}
}

@article{bozkir2023eye,
  title={Eye-tracked virtual reality: a comprehensive survey on methods and privacy challenges},
  author={Bozkir, Efe and {\"O}zdel, S{\"u}leyman and Wang, Mengdi and David-John, Brendan and Gao, Hong and Butler, Kevin and Jain, Eakta and Kasneci, Enkelejda},
  journal={Proceedings of the IEEE},
  year={2026},
  doi={10.1109/JPROC.2026.3653661}
}

@article{kasneci2024introduction,
  title={Introduction to Eye Tracking: A Hands-On Tutorial for Students and Practitioners},
  author={Kasneci, Enkelejda and Gao, Hong and Ozdel, Suleyman and Maquiling, Virmarie and Thaqi, Enkeleda and Lau, Carrie and Rong, Yao and Kasneci, Gjergji and Bozkir, Efe},
  journal={arXiv preprint arXiv:2404.15435},
  year={2024}
}

@inproceedings{bozkir2024embedding,
  title={Embedding large language models into extended reality: Opportunities and challenges for inclusion, engagement, and privacy},
  author={Bozkir, Efe and {\"O}zdel, S{\"u}leyman and Lau, Ka Hei Carrie and Wang, Mengdi and Gao, Hong and Kasneci, Enkelejda},
  booktitle={Proceedings of the 6th ACM Conference on Conversational User Interfaces},
  year={2024}
}

@article{savitzky1964smoothing,
  title={Smoothing and differentiation of data by simplified least squares procedures.},
  author={Savitzky, Abraham and Golay, Marcel JE},
  journal={Analytical chemistry},
  volume={36},
  number={8},
  year={1964},
  publisher={ACS Publications}
}

@article{mathot2018safe,
  title={Safe and sensible preprocessing and baseline correction of pupil-size data},
  author={Math{\^o}t, Sebastiaan and Fabius, Jasper and Van Heusden, Elle and Van der Stigchel, Stefan},
  journal={Behavior research methods},
  volume={50},
  year={2018},
  publisher={Springer}
}

@inproceedings{bozkir2019person,
  title={Person independent, privacy preserving, and real time assessment of cognitive load using eye tracking in a virtual reality setup},
  author={Bozkir, Efe and Geisler, David and Kasneci, Enkelejda},
  booktitle={2019 IEEE conference on virtual reality and 3D user interfaces (VR)},
  year={2019},
  organization={IEEE}
}

@inproceedings{bozkir2019assessment,
  title={Assessment of driver attention during a safety critical situation in VR to generate VR-based training},
  author={Bozkir, Efe and Geisler, David and Kasneci, Enkelejda},
  booktitle={ACM Symposium on Applied Perception 2019},
  year={2019}
}

@inproceedings{agtzidis2019360,
  title={360-degree video gaze behaviour: A ground-truth data set and a classification algorithm for eye movements},
  author={Agtzidis, Ioannis and Startsev, Mikhail and Dorr, Michael},
  booktitle={Proceedings of the 27th ACM international conference on multimedia},
  year={2019}
}

@article {Ayhan_etal_2025,
author = {Ayhan, Murat S. and Ong, Ariel Y. and Ruffell, Eden and Wagner, Siegfried K. and Merle, David A. and Keane, Pearse A.},
title = {In-context learning for data-efficient classification of diabetic retinopathy with multimodal foundation models},
year = {2025},
doi = {10.1101/2025.03.09.25323618},
publisher = {Cold Spring Harbor Laboratory Press},
journal = {medRxiv}
}

@InProceedings{hou_etal_2024,
author="Hou, Ruikun
and F{\"u}tterer, Tim
and B{\"u}hler, Babette
and Bozkir, Efe
and Gerjets, Peter
and Trautwein, Ulrich
and Kasneci, Enkelejda",
title="Automated Assessment of Encouragement and Warmth in Classrooms Leveraging Multimodal Emotional Features and ChatGPT",
booktitle="Artificial Intelligence in Education",
year="2024",
publisher="Springer Nature"
}

@misc{lau2024wrappedanansiswebunweaving,
      title={Wrapped in Anansi's Web: Unweaving the Impacts of Generative-AI Personalization and VR Immersion in Oral Storytelling}, 
      author={Ka Hei Carrie Lau and Bhada Yun and Samuel Saruba and Efe Bozkir and Enkelejda Kasneci},
      year={2024},
      doi={10.48550/arXiv.2409.16894}, 
}

@inproceedings{de2024llmr,
  title={{LLMR}: Real-time prompting of interactive worlds using large language models},
  author={De La Torre, Fernanda and Fang, Cathy Mengying and Huang, Han and Banburski-Fahey, Andrzej and Amores Fernandez, Judith and Lanier, Jaron},
  booktitle={Proceedings of the CHI Conference on Human Factors in Computing Systems},
  year={2024},
  doi = {10.1145/3613904.3642579},
  publisher = {ACM}
}

@InProceedings{farinazzo2016usability,
author="Farinazzo Martins, Val{\'e}ria
and Sampaio, Paulo N. M.
and da S. Mendes, Fernanda
and Santos Lima, Andr{\'e}
and Paiva Guimar{\~a}es, Marcelo",
title="Usability and Functionality Assessment of an Oculus Rift in Immersive and Interactive Systems Using Voice Commands",
booktitle="Virtual, Augmented and Mixed Reality",
year="2016",
}

@inproceedings{gelsomini2020embodied,
  title={Embodied learning in immersive smart spaces},
  author={Gelsomini, Mirko and Leonardi, Giulia and Garzotto, Franca},
  booktitle={Proceedings of the 2020 CHI conference on human factors in computing systems},
  year={2020}
}

@inproceedings{ciftci2017partially,
  title={Partially occluded facial action recognition and interaction in virtual reality applications},
  author={Ciftci, Umur Aybars and Zhang, Xing and Yin, Lijun},
  booktitle={2017 IEEE International Conference on Multimedia and Expo (ICME)},
  year={2017},
  organization={IEEE}
}

@inproceedings{ma2018combining,
  title={Combining brain-computer interface and eye tracking for high-speed text entry in virtual reality},
  author={Ma, Xinyao and Yao, Zhaolin and Wang, Yijun and Pei, Weihua and Chen, Hongda},
  booktitle={Proceedings of the 23rd International Conference on Intelligent User Interfaces},
  year={2018}
}

@inproceedings{radford2023robust,
  title={Robust speech recognition via large-scale weak supervision},
  author={Radford, Alec and Kim, Jong Wook and Xu, Tao and Brockman, Greg and McLeavey, Christine and Sutskever, Ilya},
  booktitle={International Conference on Machine Learning},
  year={2023},
}

@inproceedings{bolt1980put,
  title={“Put-that-there” Voice and gesture at the graphics interface},
  author={Bolt, Richard A},
  booktitle={Proceedings of the 7th annual conference on Computer graphics and interactive techniques},
  pages={262--270},
  year={1980}
}

@inproceedings{narbayev2025exploring,
  title={Exploring Pointing and Confirmation Techniques for Teleportation Across Varying Elevations in Virtual Reality},
  author={Narbayev, Bakdauren and Ullah, AKM Amanat and Sin, Jaisie and Lasserre, Patricia and Hasan, Khalad},
  booktitle={2025 IEEE International Symposium on Mixed and Augmented Reality (ISMAR)},
  pages={1671--1681},
  year={2025},
  organization={IEEE}
}

@inproceedings{2025.EDM.long-papers.9,
 address = {Palermo, Italy},
 author = {Süleyman Özdel and Can Sarpkaya and Efe Bozkir and Hong Gao and Enkelejda Kasneci},
 booktitle = {Proceedings of the 18th International Conference on Educational Data Mining},
 doi = {10.5281/zenodo.15870207},
 editor = {Caitlin Mills and Giora Alexandron and Davide Taibi and Giosuè Lo Bosco and Luc Paquette},
 isbn = {978-1-7336736-6-2},
 month = {July},
 pages = {155--169},
 publisher = {International Educational Data Mining Society},
 title = {Examining the Role of LLM-Driven Interactions on Attention and Cognitive Engagement in Virtual Classrooms},
 year = {2025}
}

@article{tatler2008systematic,
  title={Systematic tendencies in scene viewing},
  author={Tatler, Benjamin W and Vincent, Benjamin T},
  journal={Journal of Eye Movement Research},
  volume={2},
  number={2},
  year={2008},
  publisher={Bern Open Publishing}
}

@article{beatty1982task,
  title={Task-evoked pupillary responses, processing load, and the structure of processing resources},
  author={Beatty, Jackson},
  journal={Psychological Bulletin},
  volume={91},
  number={2},
  pages={276--292},
  year={1982},
  publisher={American Psychological Association}
}

@article{shiferaw2019review,
  title={A review of gaze entropy as a measure of visual scanning efficiency},
  author={Shiferaw, Brook and Downey, Luke and Crewther, David},
  journal={Neuroscience \& Biobehavioral Reviews},
  volume={96},
  pages={353--366},
  year={2019},
  publisher={Elsevier}
}

@article{payne1968workload,
  title={Percentage of pupillary dilation as a measure of item difficulty},
  author={Payne, Donald T. and Parry, Marvin E. and Harasymiw, Stephan J.},
  journal={Perception \& Psychophysics},
  volume={4},
  number={3},
  pages={139--143},
  year={1968},
  publisher={Springer}
}

@article{bahill1975main,
  title={The main sequence, a tool for studying human eye movements},
  author={Bahill, A Terry and Clark, Michael R and Stark, Lawrence},
  journal={Mathematical biosciences},
  volume={24},
  number={3-4},
  pages={191--204},
  year={1975},
  publisher={Elsevier}
}

@inproceedings{skyba2024towards,
  title={Towards natural language understanding for intuitive interactions in XR using large language models},
  author={Skyba, Kevin and Pfeiffer, Thies},
  booktitle={GI VR/AR Workshop},
  year={2024},
  organization={Gesellschaft f{\"u}r Informatik eV}
}

@inproceedings{freitag2016automatic,
  title={Automatic speed adjustment for travel through immersive virtual environments based on viewpoint quality},
  author={Freitag, Sebastian and Weyers, Benjamin and Kuhlen, Torsten W},
  booktitle={2016 IEEE Symposium on 3D User Interfaces (3DUI)},
  pages={67--70},
  year={2016},
  organization={IEEE}
}

@inproceedings{prithul2022evaluation,
  title={Evaluation of Hands-free Teleportation in VR},
  author={Prithul, Aniruddha and Bhandari, Jiwan and Spurgeon, Walker and Folmer, Eelke},
  booktitle={Proceedings of the 2022 ACM symposium on spatial user interaction},
  pages={1--6},
  year={2022}
}

@article{bangor2009determining,
  title={Determining what individual SUS scores mean: Adding an adjective rating scale},
  author={Bangor, Aaron and Kortum, Philip and Miller, James},
  journal={Journal of usability studies},
  volume={4},
  number={3},
  pages={114--123},
  year={2009},
  publisher={Usability Professionals' Association Bloomingdale, IL}
}

@article{hoerl1970ridge,
  title={Ridge regression: Biased estimation for nonorthogonal problems},
  author={Hoerl, Arthur E and Kennard, Robert W},
  journal={Technometrics},
  volume={12},
  number={1},
  pages={55--67},
  year={1970},
  publisher={Taylor \& Francis}
}

@article{cortes1995support,
  title={Support-vector networks},
  author={Cortes, Corinna and Vapnik, Vladimir},
  journal={Machine learning},
  volume={20},
  number={3},
  pages={273--297},
  year={1995},
  publisher={Springer}
}

@article{scikit-learn,
  title={Scikit-learn: Machine Learning in {P}ython},
  author={Pedregosa, F. and Varoquaux, G. and Gramfort, A. and Michel, V.
          and Thirion, B. and Grisel, O. and Blondel, M. and Prettenhofer, P.
          and Weiss, R. and Dubourg, V. and Vanderplas, J. and Passos, A. and
          Cournapeau, D. and Brucher, M. and Perrot, M. and Duchesnay, E.},
  journal={Journal of Machine Learning Research},
  volume={12},
  pages={2825--2830},
  year={2011}
}

@inproceedings{abdrabou2025gaze,
  title={From Gaze to Data: Privacy and Societal Challenges of Using Eye-tracking Data to Inform GenAI Models},
  author={Abdrabou, Yasmeen and {\"O}zdel, S{\"u}leyman and Maquiling, Virmarie and Bozkir, Efe and Kasneci, Enkelejda},
  booktitle={Proceedings of the 2025 Symposium on Eye Tracking Research and Applications},
  year={2025},
  doi={10.1145/3715669.3726788}
}

\appendix
\newpage
\section{Technical Details}
\label{sec:appendix}
\subsection{Context-Aware Prompt Construction}
\label{sec:prompt}
The core of LLM-driven locomotion system relies on a dynamic prompt structure that combines static system instructions with real-time environmental data. The prompt sent to the LLM is constructed in three distinct blocks:

\begin{enumerate}
    \item \textbf{System Instruction Block:} Is given as system prompt:\\
    You are my assistant in Unity to navigate me. You will take the inputs about the environment, then you will only give me the target position. In this task, I will ask you where I want to go. Always consider my current location, forward direction, and left direction. The forward direction is where I am looking. The left direction is my left side. The right direction is my right side. Please always consider those directions when I ask something (I will provide you with the environment information). Calculate directions based on my direction vector. Then, you will also get all the objects in the environment which I see. You have to return only the coordinates, and you cannot say anything else even if I ask. Always only return the coordinates. I will give examples for you. There are also some rules. Please follow them. You will find the best position in the environment according to my request. Always consider I am using speech-to-text, so the text input could be problematic. Try to understand it correctly in the context of our environment and task and always answer according to your understanding. Then give the coordinates in the environment according to my request.

    Teleport the player along the direction of the available road for their requests. If there are multiple objects of the same color, prioritize the object that is farthest away. Provide the output strictly in the format: {'x': x, 'y': y, 'z': z} where x, y, z are integer. 

    Notes:
    \begin{itemize}
        \item For objects of the same color, select the one farthest from the player’s position.
        \item Provide the resulting position as a JSON-like format {'x': x, 'y': y, 'z': z}.
        \item If there is no object as requested by the player, don't teleport it and return output to same inital location as the same format: {'x': x, 'y': y, 'z': z}.
        \item If there are multiple objects but the color or other specs are not specified, go to the object that is at middle distance.
    \end{itemize}

    Example:\\
    Input:\\
    Environment: I am on the \{'x': 0, 'y': 0, 'z': 0\}, my forward direction is \{'x': 0, 'y': 0, 'z': 1\}, my left direction is \{'x': -1, 'y': 0, 'z': 0\}, my right direction is \{'x': 1, 'y': 0, 'z': 0\}, and I see House\_Black => \{'x': 22, 'y': 5, 'z': 64\},House\_Orange => \{'x': 23, 'y': 5, 'z': 38\},House\_Green => \{'x': 26, 'y': 4, 'z': 15\},Four Way Crossroad => \{'x': 3, 'y': 0, 'z': 89\},House\_Gray => \{'x': -20, 'y': 4, 'z': 70\},House\_Purple => \{'x': -38, 'y': 5, 'z': 69\},House\_Red => \{'x': -20, 'y': 4, 'z': 18\},Bus stop => \{'x': 9, 'y': 0, 'z': 30\},Street Seller Stand => \{'x': -8, 'y': 0, 'z': 46\},Traffic\_light\_2 => \{'x': 8, 'y': 0, 'z': 99\},House\_Yellow => \{'x': -20, 'y': 5, 'z': 44\},Traffic\_light\_3 => \{'x': -8, 'y': 0, 'z': 85\},Traffic\_light\_3 => \{'x': -8, 'y': 0, 'z': 99\},Mailbox => \{'x': 8, 'y': 0, 'z': 43\},Black Car => \{'x': -5, 'y': 0, 'z': 62\},Traffic\_light\_2 => \{'x': 8, 'y': 0, 'z': 84\}. And available streets are \{'x': 0, 'y': 0, 'z': -2\} to \{'x': 0, 'y': 0, 'z': 98\}\\
    User command: Go to the black car\\
    Output:
    \{'x': 0.0, 'y': 0.0, 'z': 62.0\}
    
    \item \textbf{Dynamic Context Block:} Created by following structure:\\
    env\_text = "I am on the position \{start\_point\}, my forward direction is \{currentDirection\}, my left direction is \{leftDirection\}, my right direction is \{rightDirection\}, and I see "\\
    env\_text += ", ".join([f"\{obj['name']\} => \{obj['position']\}" for obj in objects])\\
    env\_text += ". And available streets are " + ", ".join([f"\{street[0]\} to \{street[1]\}" for street in availableStreets])
    \item \textbf{User Query Block:} The transcribed text from the Whisper model is appended at the end.\\
    E.g. User command: go ahead \revised{50} meters
\end{enumerate}

\subsection{Scene Extraction Algorithm}
\label{sec:scene}
The extraction of contextual scene data is handled by a custom C\# script attached to the player controller. When the record button is released, the script performs the following operations:
\begin{enumerate}
    \item \textbf{Initial Setup:} GameObjects are named based on their color and type. For example, blue car, red house, ambulance. Each GameObject is assigned to the navigation tag.
    \item \textbf{Object Retrieval:} It iterates through close \revised{(within 50\,m)} and visible active GameObjects with the pre-defined navigation tags.
    \item \textbf{Prompt Assembly:} The script concatenates the System Instruction, the serialized Scene Data, and the User Transcript into a single string. Then that string is sent via API to ChatGPT-4o.
\end{enumerate}

\subsection{Locomotion Execution Logic}
\label{sec:exec}
Upon receiving the response from the LLM, the system executes a multi-step process to ensure the destination is valid and the transition is comfortable:
\begin{itemize}
    \item \textbf{Parsing:} A regular expression extractor identifies numerical coordinate pairs within the LLM response, discarding any conversational filler text.
    \item \textbf{Street Mapping and Fallback Mechanism:} Raw coordinates may occasionally fall within non-walkable geometry (e.g., inside buildings or vehicles). The system projects such points to the nearest street centerline. In cases where the target is approximately equidistant between two streets (e.g., at an intersection), the system resolves the ambiguity by selecting the street that aligns most closely with the user's current head orientation. If the LLM response \revised{does not include a meaningful response} or parsing fails, the user \revised{stays} in the same position and the system \revised{waits} for the next request.
    \item \textbf{Visual Feedback and Transition:} Once the target is calculated, a curved ray, visually identical to standard teleportation arcs, is rendered connecting the user to the destination. This visual feedback persists for two seconds to allow the user to anticipate the movement before the instantaneous teleportation occurs.
\end{itemize}
\newpage
\subsection{Window Size Ablation and Model Confusion Matrix}
\label{app:model}

\begin{figure*}[ht]
\centering
\includegraphics[width=\textwidth]{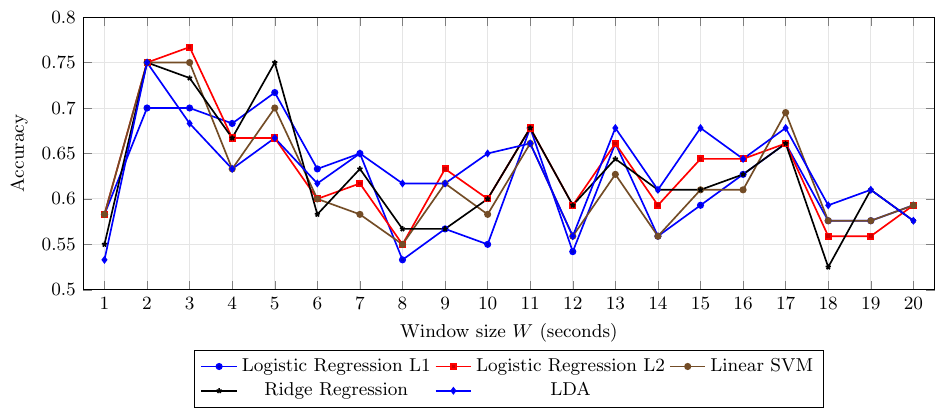}
\caption{\revised{Accuracy across window sizes ($W=1$--$20$ seconds) under leave-one-participant-out cross-validation, shown as a line chart (converted from the previous table per reviewer request). Accuracy peaks at a 3-second window (0.767 for Logistic Regression L2).}}
\revised{\Description{Line chart with window size from 1 to 20 seconds on the horizontal axis and cross-validated accuracy on the vertical axis. Each of the five models (Logistic Regression L1, Logistic Regression L2, Linear SVM, Ridge Regression, and LDA) is drawn as a separate line. Accuracy peaks around a 3-second window at roughly 0.75 to 0.77 and generally decreases for larger windows toward 0.55 to 0.66.}}
\label{tab:window_ablation_acc}
\end{figure*}
\begin{table}[ht]
\small
\centering
\caption{Confusion matrix and class-wise performance for the best-performing Logistic Regression L2 classifier.}
\label{tab:confusion_matrix}
\revised{\Description{Three-by-three confusion matrix for the best Logistic Regression L2 classifier predicting the locomotion condition. Rows are the actual class and columns the predicted class. Teleportation: 13 correct, 3 predicted as Fixed-Command, 4 as Voice-LLM (recall 65 percent). Fixed-Command: 18 correct, 2 as Voice-LLM, 0 as Teleportation (recall 90 percent). Voice-LLM: 15 correct, 3 as Teleportation, 2 as Fixed-Command (recall 75 percent). Per-class precision is 81, 78, and 71 percent; overall accuracy is 76.7 percent.}}
\revised{\begin{tabular}{@{}c l ccc c@{}}
\toprule
 & & \multicolumn{3}{c}{\textbf{Predicted}} & \\
\cmidrule(lr){3-5}
 & & \textbf{Tele.} & \textbf{Fix-cmd} & \textbf{Voice-LLM} & \textbf{Recall} \\
\midrule
\multirow{3}{*}{\rotatebox[origin=c]{90}{\textbf{Actual}}}
 & Teleportation & \textbf{13} & 3 & 4 & 65\% \\
 & Fix-command   & 0 & \textbf{18} & 2 & 90\% \\
 & Voice-LLM     & 3 & 2 & \textbf{15} & 75\% \\
\cmidrule(l){2-6}
 & Precision & 81\% & 78\% & 71\% & \textbf{Acc = 76.7\%} \\
\bottomrule
\end{tabular}}
\end{table}


\end{document}